\documentclass[pdflatex,sn-mathphys-num]{sn-jnl}% Math and Physical Sciences Numbered Reference Style
\usepackage{graphicx}%
\usepackage{multirow}%
\usepackage{amsmath,amssymb,amsfonts}%
\usepackage{amsthm}%
\usepackage{mathrsfs}%
\usepackage[title]{appendix}%
\usepackage{xcolor}%
\usepackage{textcomp}%
\usepackage{manyfoot}%
\usepackage{booktabs}%
\usepackage{algorithm}%
\usepackage{algorithmicx}%
\usepackage{algpseudocode}%
\usepackage{listings}%
\usepackage{tikz}
\usepackage{siunitx}

\newcommand{\Reynolds}{R\hspace{-0.1em}e}

\begin{document}

\title[Quick starch guide]{Quick starch guide: A perspective on shear thickening in dense non-Brownian suspensions}


\author*[1]{\fnm{Cécile} \sur{Clavaud}}\email{cecile.clavaud@univ-rennes.fr}
\equalcont{These authors contributed equally to this work.}

\author*[2]{\fnm{Abhinendra} \sur{Singh}}\email{abhinendra.singh@case.edu}
\equalcont{These authors contributed equally to this work.}

\affil[1]{\orgname{Univ Rennes, CNRS, IPR (Institut de Physique de Rennes) - UMR 6251}, \postcode{F-35000 Rennes}, \country{France}}

\affil*[2]{\orgdiv{Department of Macromolecular Science and Engineering},
\orgname{Case Western Reserve University}, \orgaddress{\city{Cleveland}, \postcode{10040}, \state{OH}, \country{USA}}}


\abstract{In this article, we provide a brief perspective on recent developments in the study of shear thickening in dense suspensions. We give a rapid overview of the state of the art and discuss current models aiming to describe this particular rheology. Although most of the experiments and simulation studies are conducted in ``ideal'' flows, where the sample is confined without an open boundary condition, we have decided to highlight more realistic flow conditions. We further provide an overview on how to relate the recently proposed constitutive models to these more practical flow conditions like pipe flow or flow down an incline. }

\keywords{shear thickening, dense suspensions, force network, network analysis, mean-field models, friction}



\maketitle

\section{Introduction}
\label{sec:introduction}

Dense suspensions are prototypical complex fluids. Their flow behavior is scientifically challenging to predict, and they are ubiquitous in various industrial, geotechnical, and biological phenomena, with examples ranging from cement transport, mudflow, and red blood cells~\cite{Denn_2014,Stickel_2005,Guazzelli_2018, Morris_2020, Jerolmack_2019,Beris_2021}. In this article, we focus on shear thickening, one of the many complex rheological behaviors they can exhibit. We will limit our discussion to suspensions of non-Brownian hard spheres suspended in a density-matched Newtonian liquid, and will therefore neglect the effects of thermal fluctuations, gravity, and particle deformability. Additionally, we will only consider viscous flows, and will thus not discuss any inertial and viscoelastic effects. Finally, as stated, we focus on shear thickening in dense suspensions. There is no clear and established definition of ``dense'' suspensions, but we can say that a suspension is dense when its particle volume fraction is ``large enough'' that its behavior is dominated by short-range interactions between particles, which are separated by distances smaller than their typical size~\cite{Ness_2022}.

Shear thickening in dense suspensions has been the subject of many reviews over the years, either directly or as one of the many existing complex rheological behaviors (see, for example,~\cite{Barnes_1989, Wagner_2009, Denn_2014} and more recently~\cite{Denn_2018, Guazzelli_2018, Singh_2023, Ness_2022, Lemaire_2023, Morris_2020}). This article is dedicated to surveying recent progress in the study of strong shear thickening observed in dense suspensions. It focuses on features encountered in realistic flows and on how to model or explain them. As it is a perspective article, it will not review all the existing literature but rather give our viewpoint on the current state of the art in the field and the promising directions for future research. It is necessarily incomplete, as we had to make choices among all the topics we would have liked to address, but we hope it will be helpful and spark new interests, ideas, and discussions. In particular, we hope to convey that shear thickening has opened up a playground for rheologists, physicists, engineers, chemists, tribologists, and mathematicians to play on.

\section{Shear thickening: what it is, and what it is not}
\label{sec:definitions}

Shear thickening (ST) is an increase in the viscosity $\eta = \tau/\dot{\gamma}$ of a complex fluid with the applied shear rate $\dot{\gamma}$ or shear stress $\tau$. In dense shear-thickening suspensions, this increase, as a function of the applied shear rate, can be either continuous, in which case the behavior is called continuous shear thickening (CST), or discontinuous, giving discontinuous shear thickening (DST). CST is usually observed at lower volume fractions $\phi$. Shear thickening becomes stronger as $\phi$ increases, until DST is observed, where $\eta$ increases abruptly by orders of magnitude over a narrow range of $\dot{\gamma}: \partial \eta/\partial \dot{\gamma} \to \infty$. The conditions ($\phi,\dot{\gamma}$) at which DST is observed depend sensitively on the particle properties, i.e., roughness, size, shape, etc., as well as on the chemical physics of both the particle and fluid phases~\cite{Singh_2019}.

\subsection{The many features of shear thickening in dense suspensions}
\label{subsec:history}

The first papers (to our knowledge) to report on this intriguing rheology are from Williamson and Heckert~\cite{Williamson_1931} and Freundlich and R\"oder~\cite{Freundlich_1938}, and it is worth discussing them a little. In both papers, the authors observe a saturation of their studied suspension's flow rate with the applied stress. Williamson and Heckert conducted an extensive study of cornstarch suspensions in various suspending fluids, while Freundlich and R\"oder studied suspensions of quartz powder, rice starch, and potato starch in water, with or without added electrolytes. In both studies, the authors note that the behavior is strongly dependent on the fluid-particles interactions: if the particles are not well-dispersed in the fluid, they do not observe the flow rate saturation. They also both postulate that contacts between the particles play an important role. Freundlich and R\"oder report that the flow rate saturation comes with a periodic flow instability that causes the suspensions to fracture. Finally, Williamson and Heckert postulate the existence of a critical stress scale in the system to explain the behavior. We will come back to the observations and ideas put forward in these two papers throughout this perspective article.

\begin{figure}[!ht]
    \begin{center}
    \includegraphics[width=\linewidth]{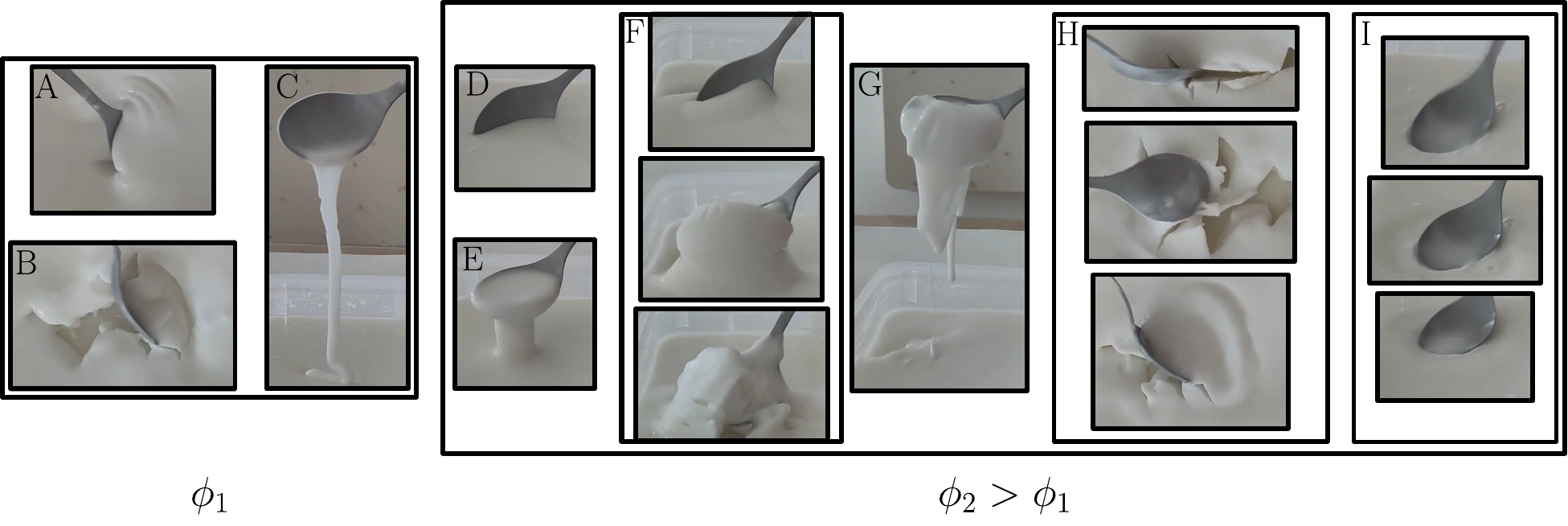}
    \end{center}
    \caption{Some of the (many) features of shear thickening in dense shear-thickening suspensions. Suspension used: commercial potato starch suspended in tap water. Left: behavior at a first packing fraction $\phi_1$ ($\phi_1 \simeq 0.55$ wt\%). A. Scooping-up of the suspension at medium speed, versus B. fast scooping-up, and C. subsequent flow of the suspension out of the spoon (irrespective of the scooping-up speed). We can see signs of wrinkling, fracture that rapidly relaxes in the wake of the spoon, and fracture in the filament flowing out of the spoon. Right: behavior at a packing fraction $\phi_2 > \phi_1$ ($\phi_2 \simeq 0.57$ wt\%). D. Slow scooping-up of the suspension and E. subsequent flow out of the spoon, for a small spoon height. The suspension here behaves like a very viscous liquid. F. Fast scooping-up of the suspension, and G. subsequent flow out of the spoon. Again we can see signs of fracture, with (slower) relaxation where the suspension is no longer under stress. H. Fracture of the suspension under large stress. I.~Solidification under impact.}
    \label{fig:features}
\end{figure}

Since these two articles, there have been quite a lot of experimental studies on the subject, revealing different features of shear-thickening behavior in dense suspensions. Metzner and Whitlock reported on the ability of such suspensions to fracture, stating that ``\emph{the fracturing of the fluid was not only visible but also audible}''~\cite{Metzner_1958}. Numerous experiments pointed to DST being associated not only with an abrupt increase in viscosity, but also with giant fluctuations in stress (or viscosity) at the onset of DST~\cite{Hoffman_1972, Hoffman_1974, Lootens_2003, Lootens_2004, Bender_1996, Boersma_1991, DHaene_1993, Xu_2020}. Saint-Michel \textit{et al.}~\cite{Saint-Michel_2018}, using cornstarch suspensions in a concentric cylinder and simultaneously imaging the flow with ultrasonic echography, observed and provided a statistical analysis of temporal fluctuations in the shear rate. They correlated it with spatiotemporal local particle dynamics and reported that the bulk suspension remains homogeneous during the DST transition, along with the existence of transient localized bands that travel along the vorticity direction. Rathee \textit{et al.}~\cite{Rathee_2017, Rathee_2020, Rathee_2022}, in a series of studies using boundary stress microscopy, showed the existence of an inhomogeneous stress field as the suspension shear thickens. Along with this, the authors also revealed the existence of high local stress regions that propagate along the flow direction of the top boundary. Along these lines, Ovarlez \textit{et al.}~\cite{Ovarlez_2020} also reported density fluctuations that appear as periodic waves in the flow direction and break the azimuthal symmetry, accompanied by high normal stress fluctuations of the same periodicity. Finally, one of the most famous aspects of shear-thickening suspensions is their response to impact and their ability to transiently support loads and dissipate energy. This has also been the focus of many papers (see, for example,~\cite{Waitukaitis_2012, Roche_2013, Han_2018, Brassard_2021, Peters_2016, James_2018, Chen_2023}), showing the development and travel of a jamming front under impact or studying the cracks formed after impact. Much of the early experimental, theoretical, and numerical work on shear-thickening suspensions has focused on first understanding the average bulk behavior in standard rheological geometries and in steady flow, but it is interesting to note that the unstable nature of the flow in the DST regime was already discussed in~\cite{Freundlich_1938}. We illustrate some of the different features of shear thickening in dense suspensions in Fig.~\ref{fig:features}.

The previous paragraph concerns features related to flow behavior. In addition to this, both  Refs.~\cite{Williamson_1931} and~\cite{Freundlich_1938} discussed the importance of fluid-particle interactions in the presence or absence of shear thickening. This point seems to have received less interest, although it is of crucial importance. Indeed, fluid-particle interactions, in addition to the physical characteristics of the particles, are responsible for setting the particle-particle interactions (or constraints, see Sec.~\ref{sec:constraints}). Some recent studies on suspensions of small ($\sim \qtyrange{0.1}{4}{\micro \meter}$) particles
explore the effect of particle-particle interactions on CST and DST, either through directly manipulating the particles' roughness~\cite{Hsu_2018, Hsu_2021, Lootens_2005, Hsiao_2017, Pradeep_2021}, changing their physical and chemical interactions~\cite{James_2018, James_2019, Kim_2024, Chen_2023}, or both~\cite{Bourrianne_2022}. In the concrete industry, the effect of superplasticizers (polyelectrolytes that can adsorb on the surface of the particles in the cement slurry) on the suspension rheology has long been documented (see, for example,~\cite{Neuville_2012, Bossis_2017}), and this effect is also studied in a more formal context (see for example~\cite{Richards_2021}). James \textit{et al.}~\cite{James_2018} emphasized the pivotal role of physiochemical interactions in the manifestation of DST in suspensions containing both Brownian and non-Brownian (cornstarch) particles. In the case of non-Brownian suspensions, Clavaud \textit{et al.} showed in~\cite{Clavaud_2017} that changing the suspending fluid could change fluid-mediated particle-particle interactions, thus changing the rheological behavior of the suspension. In line with the results from~\cite{Williamson_1931}, Oyarte G\'alvez \textit{et al.}~\cite{OyarteGalvez_2017} showed that the nature of the suspending fluid of cornstarch suspensions had a dramatic effect on the suspension rheology. With all this in mind, we wanted to highlight the specific case of suspensions of cornstarch in water, often taken to be the prototypical example of non-Brownian shear-thickening suspensions. Gauthier \textit{et al.} showed in~\cite{Gauthier_2023} that they exhibit signs of adhesion, and that their behavior is not always the same as that of other shear-thickening suspensions. In light of these observations, we felt it was important to mention the critical importance of the fluid-particle interactions, and remind the reader to be careful before assuming the generality of an observed behavior.

\subsection{Constitutive laws for dense suspensions in simple shear and steady-state: why shear thickening is unexpected}
\label{subsec:suspension_rheology_and_dilatancy}

\begin{figure}[!ht]
    \begin{center}
    \includegraphics[width=\linewidth]{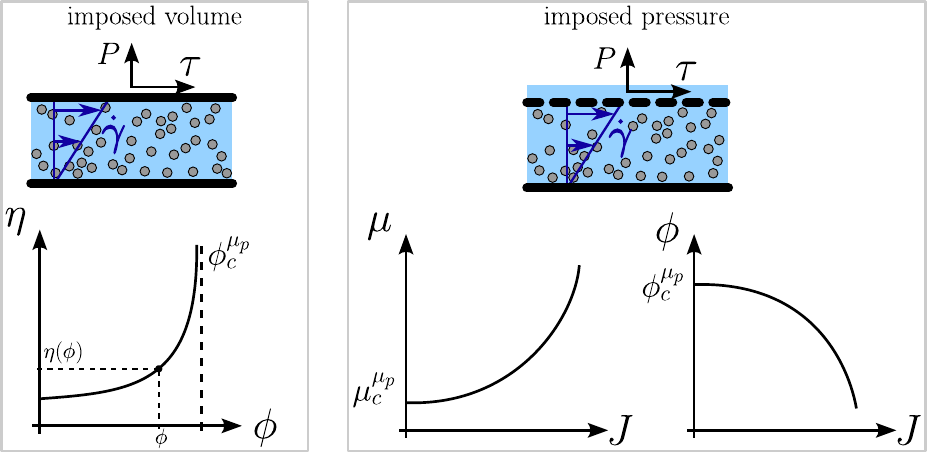}
    \end{center}
    \caption{Steady-state simple shear rheology of a suspension of non-Brownian hard spheres in a density matched Newtonian liquid, undergoing a viscous flow. Left: volume-imposed case. The suspension's viscosity $\eta$ is a function of only one parameter, its particle volume fraction $\phi$: $\eta(\phi) = \eta_f \eta_s(\phi)$. Right: pressure-imposed case. A certain pressure $P$ is applied only to the particulate phase of the suspension, as represented here with a grid that allows the fluid to pass through.}
    \label{fig:simple_shear_rheology}
\end{figure}

In simple shear and in the steady state, and in the framework of this paper, dimensional analysis dictates the following constitutive laws, which have been experimentally confirmed (see~\cite{Guazzelli_2018} and references therein):
\begin{align}
\label{eq:constitutive_laws_standard_suspensions}
&\left\{
\begin{aligned}
\tau &= \eta_f\dot{\gamma} \eta_s(\phi)\\
P &= \eta_f\dot{\gamma} \eta_n(\phi)
\end{aligned}
\right.
& & \text{or}
& &\left\{
\begin{aligned}
\tau &= \mu(J)P\\
\phi &= \phi(J)
\end{aligned}
\right.
\end{align}
in, respectively, a volume-imposed or a pressure-imposed geometry. Here,~$\eta_f$ is the suspending fluid's viscosity and~$\dot{\gamma}$ is the imposed shear rate. In a volume-imposed geometry, the particle volume fraction~$\phi$ is imposed, and the shear stress~$\tau$ and particle pressure~$P$ depend on~$\phi$ through the functions~$\eta_s$ and~$\eta_n$, which diverge at the jamming packing fraction~$\phi_c^{\mu_p}$ above which the suspension cannot flow. This jamming packing fraction depends on the microscopic friction coefficient~$\mu_p$, that is, the particle-particle friction coefficient, and we have~$\phi_c^{\mu_p \neq 0}<\phi_c^{\mu_p = 0}$. In a pressure-imposed geometry, the particle pressure~$P$ is imposed, and both~$\phi$ and the ratio~$\tau/P$ are controlled by the viscous number~$J = \eta_f\dot{\gamma}/P$ (note that $J$ is the inverse of the normal viscosity $\eta_n$ as defined in an imposed volume geometry). By analogy to Coulomb's law of friction, the function~$\mu(J)$ is called the macroscopic friction coefficient of the suspension. We have $\phi_c^{\mu_p} = \lim\limits_{J \to 0} \phi(J)$, and~$\mu$ has a minimum value~$\mu_c^{\mu_p}$ at low~$J$, which also depends on~$\mu_p$. In the absence of any additional ingredient, as summarized in Fig.~\ref{fig:simple_shear_rheology}, the rheology of such a suspension is thus \emph{quasi}-Newtonian: its viscosity is independent of the shear rate. The existence of shear thickening points to a missing parameter in the description of shear-thickening suspensions. We will come back to this in Section~\ref{sec:WCmodel}.

Note that what we have presented here is a simplified view of the rheology of dense suspensions, for two main reasons. First, the definition of $\phi_c^{\mu_p\neq 0}$ is experimental, mostly based on a fit to experimental data, and hides a more complex phenomenology. There are, \textit{a priori}, three critical packing fractions in dense suspenions: $\phi^{\mu_p\neq 0}_\text{min. rig.}$ at which minimally rigid frictional clusters start to appear in the suspension (see Sec.~\ref{sec:beyond_mean_field}), $\phi^{\mu_p\neq 0}_\text{SJ}$ above which the suspension can only flow until it accumulates a certain amount of strain, which creates an anisotropic frictional microstructure preventing flow in the strain direction (shear-induced jamming, see Sec.~\ref{sec:WCmodel} and~\ref{sec:beyond_mean_field}), and $\phi^{\mu_p\neq 0}_\text{IJ}$, the packing fraction above which the system is isotropically jammed. (All three \textit{a priori} depend on the particle--particle friction coefficient, and we postulate that when $\mu_p = 0$, they are all equal). Disentangling these is a complicated task, and is the focus of some recent studies~\cite{Seto_2019, Naald_2024}. Although we will discuss rigidity and shear-jamming, we will still restrict ourselves to a formulation in which there is only one critical packing fraction: $\phi_c^{\mu_p\neq 0}$. Second, these constitutive laws do not fully reflect the rheology of dense suspensions. They do not predict, for example, the existence of normal stress differences, hence the \emph{quasi}-Newtonian appellation. Additionally, they only describe steady-state behavior and do not consider the transients leading to it. Finally, the relations are written between the averaged values of the considered physical quantities, again not allowing for the description of any transient phenomenon rising from fluctuations in the system~\cite{Guazzelli_2018}, nor for the consideration of any nonlocal effect, which have been shown to be crucial in explaining the quasi-static response of other particulate systems~(see~\cite{Kamrin_2024} for a review, and see references therein). Coming back to the definition of a dense suspension, we can rephrase it as: the suspension's particle volume fraction is ``close enough'' to its jamming value that it leads to significant non-Newtonian behavior, even for suspensions that have a mostly rate-independent viscosity.

\subsection{Dilatancy, dilatancy, and shear thickening}
\label{subsec:dilatancy}

From the beginning, shear thickening has had a troubled relationship with dilatancy. Willamson and Heckert argued in~\cite{Williamson_1931} that it was necessarily distinct from dilatancy, in response to which Freundlich and R\"oder stated in~\cite{Freundlich_1938} that it was the same phenomenon. Let us state here clearly that shear thickening and dilatancy are, indeed, two different phenomena. Dilatancy, which we discuss in this section, is a universal feature of all frictional particulate systems (both dry granular matter and suspensions), whereas shear thickening is not.

\begin{figure}[!ht]
\begin{center}
\includegraphics[width=\linewidth]{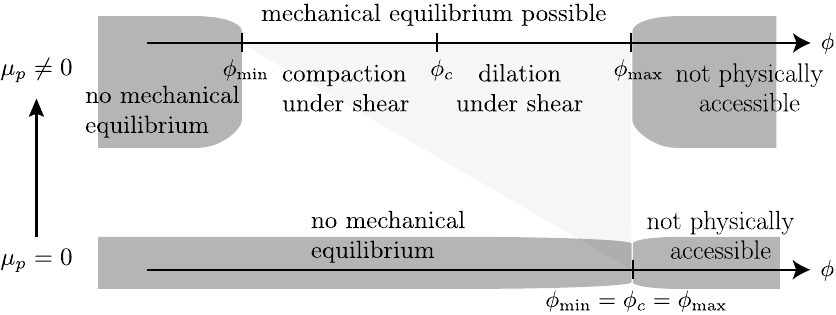}
\caption{Range of packing fractions for which mechanical equilibrium under gravity is possible for a granular pile, as a function of the microscopic solid friction coefficient~$\mu_p$.}
\label{figure_comp_dil_phi}
\end{center}
\end{figure}

Frictional granular piles (wether immersed or not) have stable mechanical states for a range of packing fractions $[\phi_\text{min}, \phi_\text{max}]$ starting below and ending above the jamming packing fraction $\phi_c^{\mu_p}$  (see~\cite{Andreotti_2013} and Fig.~\ref{figure_comp_dil_phi}, top). This has important consequences for their transient behavior under deformation, as it leads to the well-known Reynolds dilatancy effect~\cite{Reynolds_1885}. When sheared at fixed pressure and in quasi-static conditions, dense granular piles with initial packing fractions $\phi > \phi_c^{\mu_p}$ cannot flow without dilating. The packing fraction thus decreases with the strain~$\gamma$ until it reaches~$\phi_c^{\mu_p}$. Conversely, if the medium is initially loose, the packing fraction increases with~$\gamma$, until it reaches~$\phi_c^{\mu_p}$. After a strain~$\gamma_0 \sim O(1)$, the system has forgotten its initial state (loose or dense), and simply flows at~$\phi_c^{\mu_p}$~\cite{Andreotti_2013, Singh_2013}. This property of being mechanically stable at different packing fractions, and of compacting or dilating under shear, is due to the existence of frictional solid contacts between the particles. In contrast, for frictionless hard spheres, the conditions of mechanical equilibrium and non-overlap between grains impose only one possible coordination number for the pile~\cite{Andreotti_2013}. This suggests that perfectly frictionless beads packs at equilibrium only have access to one packing fraction, and that this packing fraction is $\phi_c^{\mu_p=0}$ (see~Fig.~\ref{figure_comp_dil_phi}, bottom), a conclusion that is supported by numerical simulations~\cite{Peyneau_2008} and experimental work~\cite{Clavaud_2017}.

Additionally, we have seen that under steady but non quasi-static flow, the packing fraction~$\phi$ decreases with the viscous number~$J$. By contrast with Reynolds dilatancy, which is a transient phenomenon, this dilatancy under flow is simply the steady-state constitutive law for the packing fraction.

Both dilatancies can be coupled when shearing a suspension with an initial packing fraction~$\phi_i \neq \phi(J)$ at a given~$J$. In this case, the packing fraction of the suspension will increase or decrease with strain until it reaches~$\phi(J)$. This transient dilation or compaction towards the steady-state value~$\phi(J)$ has important consequences on the behavior of suspensions after a change in the flow conditions. Indeed, it must be accompanied by an inward or outward liquid flow, and thus a positive or negative pore pressure between the grains. The suspension's behavior shortly after a change in~$J$ is then set by this transient flow and its interaction with the grains, rather than the steady-state suspension's rheology~\cite{Athani_2022}.

\section{The frictional transition scenario and the Wyart and Cates model}
\label{sec:WCmodel}

As discussed in Subsec.~\ref{subsec:suspension_rheology_and_dilatancy}, dimensional analysis requires that an additional ingredient, namely, a stress scale, is added to the parameters describing the suspension so that its viscosity can depend on the shear stress (or rate). A key idea put forward by Seto \textit{et al.}~\cite{Seto_2013a} to explain shear thickening in dense suspensions was the occurrence of frictional contacts between particles above a certain critical force. This frictionless-to-frictional transition led to a paradigm shift in understanding the physics behind DST, and is the core of this perspective.

\begin{figure}[!ht]
\begin{center}
\includegraphics[width=\linewidth]{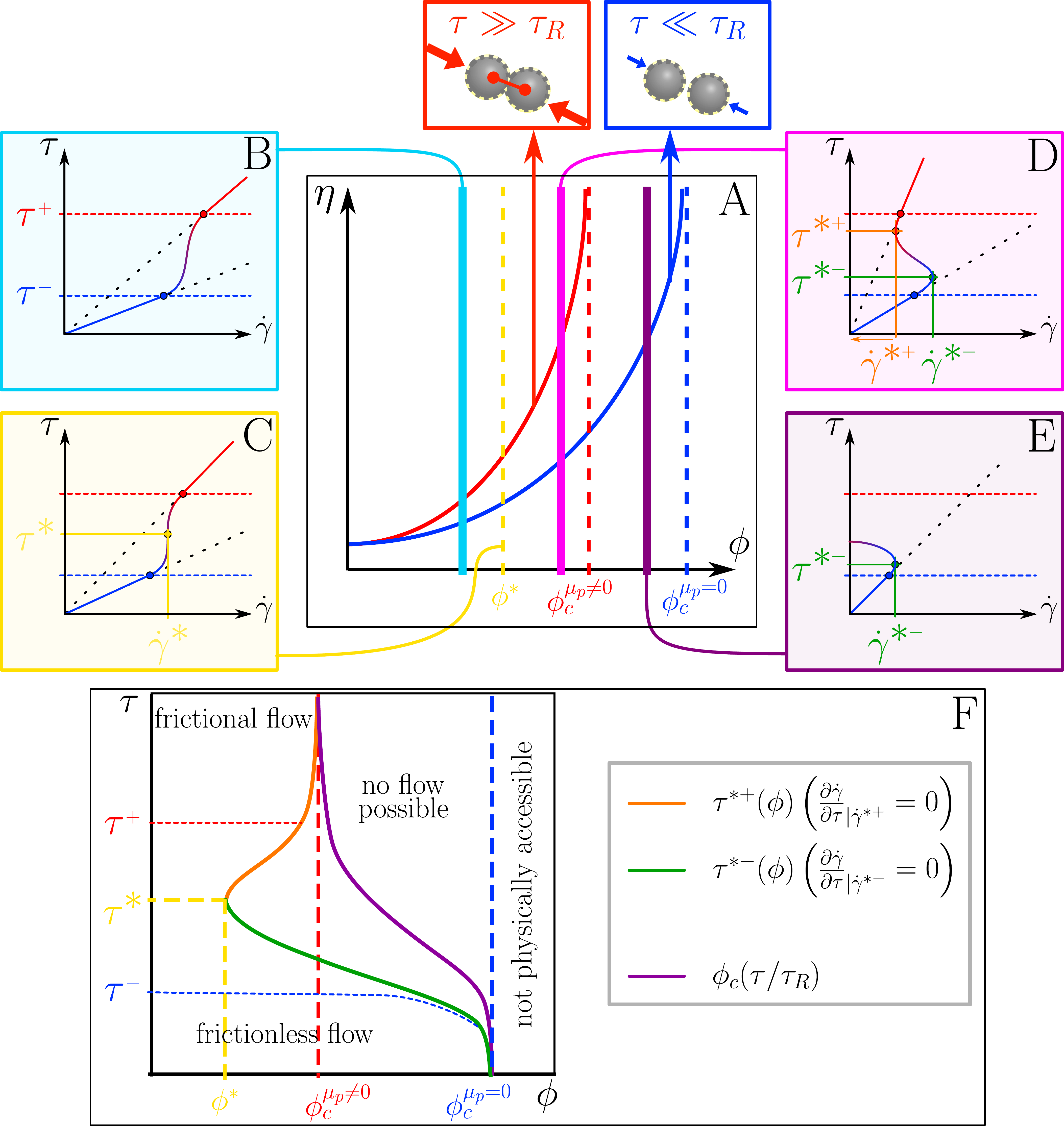}
\end{center}
\caption{Shear-thickening behavior in a dense suspension, in the framework of the Wyart and Cates model. A. The two accessible rheological branches. In red, the frictional branch, diverging at $\phi_c^{\mu_p\neq 0}$, and in blue, the frictionless branch, diverging at $\phi_c^{\mu_p = 0}$. B, C, D and E: the different flow curves resulting from the transition from the frictionless to the frictional branch, at different packing fractions. B: CST at low values of the packing fraction (monotonic $\tau(\dot{\gamma})$ relation). C: onset of DST at a packing fraction $\phi^{*}$, with the apparition of a vertical tangent in the flow curve. D: for $\phi^{*}<\phi<\phi_c^{\mu_p\neq 0}$, DST between the frictionless and the frictional branches (non-monotonic $\tau(\dot{\gamma})$ relation). Shear stresses $\tau^{*-}$ and $\tau^{*+}$: locations of the vertical tangents for $\tau\left(\dot{\gamma}\right)$ (equivalently, locations of $\frac{\partial \dot{\gamma}}{\partial \tau} = 0$). E: for $\phi>\phi_c^{\mu_p\neq 0}$, DST going from the frictionless branch to the shear-jammed frictional state. F: resulting phase diagram. For $\tau \gg \tau_R$, the suspension is in a frictional state, and for $\tau \ll \tau_R$, it is in an effective frictionless state, thus undergoing a frictionless flow. Above $\phi_c\left(\tau/\tau_R\right)$ (which depends on the shear stress $\tau$), no flow is possible.}
\label{fig:WC_Model}
\end{figure}

The main idea on which this series of works has been focused is the \emph{stress-driven activation of interparticle friction}. The authors suggest that the minimal model for shear thickening (especially DST) should consider only three forces: a short-range repulsive force $F_R$, the lubrication force $F_H$, and frictional contact force $F_C$. They assume the breakdown of the lubrication film, as identified in a previous study by Ball and Melrose~\cite{Ball_1995}, which is of fundamental importance for the formation of contacts between particles. Since lubrication breakdown occurs in Stokes flow, it is assumed to be rate-independent~\cite{Ball_1995, Morris_2020}. This, along with experimental evidence that roughness drastically affects shear-thickening behavior, motivated simulation studies to include frictional contacts~\cite{Seto_2013a, Mari_2014}. The resulting frictional transition scenario is as follows. The short-range repulsive force $F_R$ leads to a critical stress $\tau_R \sim F_R/\pi a^2$, which must be overcome for the particles to form frictional contacts (where $a$ is the particle size). Below this critical stress $\tau_R$, external shear is balanced by the repulsive force, which keeps particle surfaces apart. As shear stress (or rate) increases, that is, for $\tau>\tau_R$, the particles start forming more and more frictional contacts. In the stress states $\tau \ll \tau_R$, the particles are in an \emph{effective} frictionless state, and the suspension's rheology evolves on the frictionless branch (see Fig.~\ref{fig:WC_Model} A), with a viscosity diverging at the frictionless jamming point $\phi_c^{\mu_p=0}$ (corresponding to random close packing in the monodisperse case). For $\tau \gg \tau_R$, almost all contacts are frictional, and the suspension evolves on the frictional branch, which has a higher viscosity (see Fig.~\ref{fig:WC_Model} A)~\cite{Seto_2013a, Mari_2014}. As stated in Subsec.~\ref{subsec:suspension_rheology_and_dilatancy}, the viscosity of the frictional branch diverges at a volume fraction $\phi_c^{\mu_p\neq 0} < \phi_c^{\mu_p = 0}$. The ability of shear-thickening suspensions to switch between a frictionless and a frictional rheological branch leads to the shear-thickening transition. We find it interesting to come back to~\cite{Williamson_1931} and~\cite{Freundlich_1938}, and point out that both suggested particle-particle contacts may play a role in shear thickening. From their observations, Williamson and Heckert also postulated the existence of a stress scale in the system that would explain the observed unusual behavior. One last thing that is critical to mention here is that friction between the particles and hydrodynamics alone will yield the rate-independent quasi-Newtonian rheology described in Subsec.~\ref{subsec:suspension_rheology_and_dilatancy}, Eq.~\eqref{eq:constitutive_laws_standard_suspensions}. The repulsive force is necessary to obtain a stress-driven transition from the frictionless to the frictional state.

This idea led to the development of a constitutive model by Wyart and Cates~\cite{Wyart_2014}, which captures both CST and DST. The authors integrated the activation of frictional contacts into the constitutive laws given in equations~\eqref{eq:constitutive_laws_standard_suspensions} in the volume-imposed context. To do that, they introduce $f$, the fraction of frictional contacts between the particles, and write it as a function of $\tau/\tau_R$, such that $f\left(\tau/\tau_R\right) = 0$ if $\tau/\tau_R \ll 1$ and increases smoothly with $\tau$ to $f\left(\tau/\tau_R\right) = 1$ for $\tau/\tau_R \gg 1$. They then write the jamming packing fraction of the suspension as an interpolation between the frictionless and the frictional ones:
\begin{equation}
\label{eq:phi_c_Pi}
\phi_c\left(\frac{\tau}{\tau_R}\right) = f\left(\frac{\tau}{\tau_R}\right)\phi_c^{\mu_p\neq 0} + \left(1-f\left(\frac{\tau}{\tau_R}\right)\right)\phi_c^{\mu_p=0}.
\end{equation}
Writing that $\eta_s$ and $\eta_n$ scale as $\left(\phi - \phi_c\right)^{-2}$ (see for example~\cite{Guazzelli_2018}), they finally get the rheological curves sketched in Fig.~\ref{fig:WC_Model} B, C, D and E. Starting from the end, let us first note that $\phi>\phi_c^{\mu_p = 0}$ is not physically accessible, as stated in Subsec.~\ref{subsec:dilatancy}. For $\phi<\phi_c^{\mu_p = 0}$, the behavior is quasi-Newtonian at stress levels below $\tau^{-}$, and for $\phi < \phi_c^{\mu_p \neq 0}$, it is also quasi Newtonian above $\tau^{+}$. With increasing $\phi$, the viscosity of the frictional branch increases faster than that of the frictionless one. As a consequence, there is a critical volume fraction $\phi^{*} < \phi_c^{\mu_p\neq 0}$ at which the curve joining the frictionless and frictional branches has a vertical tangent. The exact value of $\phi^{*}$, and of the corresponding critical shear stress $\tau^{*}$, depends on the functional form chosen for $f$. The same goes for the values of $\tau^{+}$ and $\tau^{-}$. Note that $\tau^{*} \neq \tau_R$ (we have $\tau^{*}\simeq 1.5\, \tau_R$, see~\cite{Wyart_2014, Singh_2018}). Below this critical volume fraction, that is, at $\phi < \phi^{*}$, the curve connecting the frictionless and the frictional states is monotonic, and the suspension undergoes CST between $\tau^{-}$ and $\tau^{+}$. Above $\phi^{*}$, the flow curve becomes S-shaped. Above $\phi^{*}$  but below $\phi_c^{\mu_p\neq 0}$, the suspension undergoes CST above $\tau^{-}$, then DST between $\tau^{*-}$ and $\tau^{*+}$, where $\frac{\partial \tau}{\partial \dot{\gamma}}<0$. The shear stresses $\tau^{*-}$ and $\tau^{*+}$ correspond to the two points at which $\tau\left(\dot{\gamma}\right)$ has a vertical tangent. They are associated with shear rates $\dot{\gamma}^{*-}$ and $\dot{\gamma}^{*+}$. Both $\dot{\gamma}^{*-}$ and $\dot{\gamma}^{*+}$ decrease with the packing fraction, with $\dot{\gamma}^{*+} \to 0$ when $\phi \to \phi_c^{\mu_p\neq 0}$ and $\dot{\gamma}^{*-} \to 0$ when $\phi \to \phi_c^{\mu_p = 0}$. Again, note that the DST onset stress $\tau^{*-}$ is different from $\tau_R$ (for one thing, $\tau^{*-}$ depends on~$\phi$). Finally, when the frictional state is jammed, that is, for $\phi>\phi_c^{\mu_p\neq 0}$, the system can only flow at low and intermediate stresses, while frictional contacts cause the suspension to shear-jam at higher stresses. In this case, the flow curve tends towards zero shear rate at high stresses.

Note that by construction, the Wyart and Cates model is a mean-field and steady-state model: it is not designed to capture any transient behavior or describe local fluctuations or heterogeneities, similarly to eq.~\eqref{eq:constitutive_laws_standard_suspensions}. The original construction by Wyart and Cates defined $f$ in terms of particle pressure~\cite{Wyart_2014}, but these forms are interchangeable, since shear and normal stresses are related by an $O(1)$ ratio. Following this, work by Singh \textit{et al.}~\cite{Singh_2018} showed that the Wyart and Cates model is able to predict the full rheological response of concentrated shear-thickening suspensions, that is: viscosity, normal stress differences, and particle pressure. The authors demonstrated that viscometric functions can be cast into functional forms that can be parameterized by the volume fraction, the applied stress relative to the repulsive force scale, and the interparticle friction coefficient.

Let us finish this section by coming back to what happens in shear-thickening suspensions when $\phi>\phi_c\left(\tau/\tau_R\right)$. In accordance with the literature, we term the crossing of this $\phi_c\left(\tau/\tau_R\right)$ line with increasing stress as shear-jamming (SJ). However, this appellation deserves a comment. Here, as long as the applied stress $\tau$ is such that $\phi>\phi_c\left(\tau/\tau_R\right)$, the shear-thickening suspension stays jammed, but it unjams if this stress is decreased or removed. If the stress is lowered to values such that $\tau<\tau^{-}$, the suspension goes back to its stable frictionless, flowable state. This is different from shear-jamming in non shear-thickening suspensions, which is caused by the creation, under shear, of a microstructure that endures even at vanishing (but finite) stresses. In this case, the shear-jammed suspension stays jammed in the direction of the previously applied stress, even at very small stresses. As we have seen in Subsec.~\ref{subsec:suspension_rheology_and_dilatancy}, non shear-thickening suspensions do not have an internal stress scale. In order to undergo shear-jamming, they need to be dense enough, and to accumulate a certain amount of strain, no matter the value of the applied stress leading to this strain. SJ for shear-thickening suspensions is shear-jamming \emph{under stress} (stress is necessary to achieve and to maintain the jammed state), while SJ for non shear-thickening suspensions is \emph{strain-induced} shear-jamming (the shear-jammed state is created by the accumulation of strain, and provided it stays finite, the stress can become very low without the shear-jammed state failing). For the sake of readability, we will call all these behaviors shear-jamming, but we invite the reader to keep in mind the differences between them. We felt it was important to clarify that SJ for shear-thickening and for non shear-thickening dense suspensions are conceptually different, despite being called the same. This similarity in naming is due, in part, to the fact that for a while, suspensions of cornstarch in water were seen as prototypical examples of dense suspensions. As we have seen, not all dense suspensions shear thicken, and cornstarch particles suspended in water are adhesive to a certain degree. Cornstarch suspensions in water are therefore neither prototypical dense suspensions, nor prototypical shear-thickening dense suspensions.

\section{Other models: what if lubrication didn't break down?}
\label{sec:hydrodynamics}

In the previous section, we assumed that particles suspended in a Newtonian fluid can come close enough to each other to form frictional contacts. As discussed, this idea is based on the assumption that the lubrication films between particles cannot thin indefinitely, and have to break up at some point. However, lubrication breakdown is not always assumed. Earlier work by Jeffrey and co-workers~\cite{Jeffrey_1984, Jeffrey_1992} led to the development of hydrodynamic interactions between a pair of particles in terms of resistance matrix and relative particle velocity. Based on this, Brady and coworkers developed a discrete particle simulation tool, called the Stokesian Dynamics approach~\cite{Bossis_1987, Brady_1988}. This approach splits the hydrodynamic interactions into close-ranged or near-field interactions (lubrication) and far-field (full hydrodynamics). With this approximation, the leading terms in the resistance matrix associated with the normal and tangential motions of the particles diverge as $1/h$ and $1/\log(h)$ respectively, $h$ being the surface separation. This method is able to well reproduce the rheology of Brownian~\cite{Morris_2002} and non-Brownian~\cite{Foss_2000} suspensions until strong shear thickening is observed experimentally. As the shear rate (or stress) is increased, the particles come closer and closer. Since the hydrodynamic stress scales as $1/h$, the dissipative forces become significantly larger than external deformation. These hydrodynamically nearly touching  particles align along the dominant principal axis, where a strong lubrication force between small gaps causes strong correlated motion. This led to the proposition of the hydrocluster mechanism by Wagner and Brady~\cite{Wagner_2009}. The hydrocluster model successfully reproduces the onset of shear-thickening and weak continuous shear thickening; but it cannot reproduce strong CST or DST.

The main challenge with the hydrodynamics approach is the basic assumption of smooth hard-sphere particles, whereas in real life smooth hard spheres do not exist. Based on this, Jamali and Brady~\cite{Jamali_2019} proposed a modified lubrication approach. Using a discrete particle simulation approach called dissipative particle dynamics, the authors simulated explicit asperities on top of smooth particles. In this case, when the particles surfaces are separated by a distance $h$, the particles asperities can be closer than $h$. This leads to asperity-asperity lubrication forces being stronger than base particle-particle lubrication forces. As the asperity size increases, stronger hydrodynamic dissipation leads to larger viscosity, eventually leading to strong CST and, more importantly, DST. Wang \textit{et al.}~\cite{Wang_2020} further showed that strengthening the tangential logarithmic divergence with a polynomial one, the terms of polynomial functions being dependent on the size of the asperity, can essentially reproduce the DST. In a related but different approach, Rosales-Romero and coworkers~\cite{Rosales-Romero_2024} consider the effect of confinement between asperities on the suspending fluid's viscosity, and show that this can also lead to shear thickening.

To conclude this section, what is clear is that idealized smooth hard spheres do not exist in real life. Simulations considering static friction as well as modified hydrodynamics (considering the effect of asperities) can reproduce the DST behavior. Simulations considering static friction also reproduce the SJ behavior observed experimentally~\cite{Peters_2016, James_2018}, since friction keeps particles in enduring contacts and will be present even in the limit of zero shear rate. However, SJ is not expected in the absence of static friction, as the fluid-mediated forces vanish in the limit of vanishing shear rate, thus restoring a finite viscosity. The origin of DST is still debated, but one thing is clear: Near-contact interactions between particles with lubrication, surface roughness, and repulsion are the minimal ingredients. Surface roughness can either lead to friction between particles~\cite{Seto_2013a, Mari_2014, Clavaud_2017, Comtet_2017, Singh_2018, Singh_2020, More_2020} or lead to enhanced lubrication between asperities~\cite{Jamali_2019, Wang_2020, Rosales-Romero_2024}.

\section{Constrained versus unconstrained particle motion: one point of view to unite them all}
\label{sec:constraints}

\begin{figure}[!ht]
\begin{center}
\includegraphics[width=\linewidth]{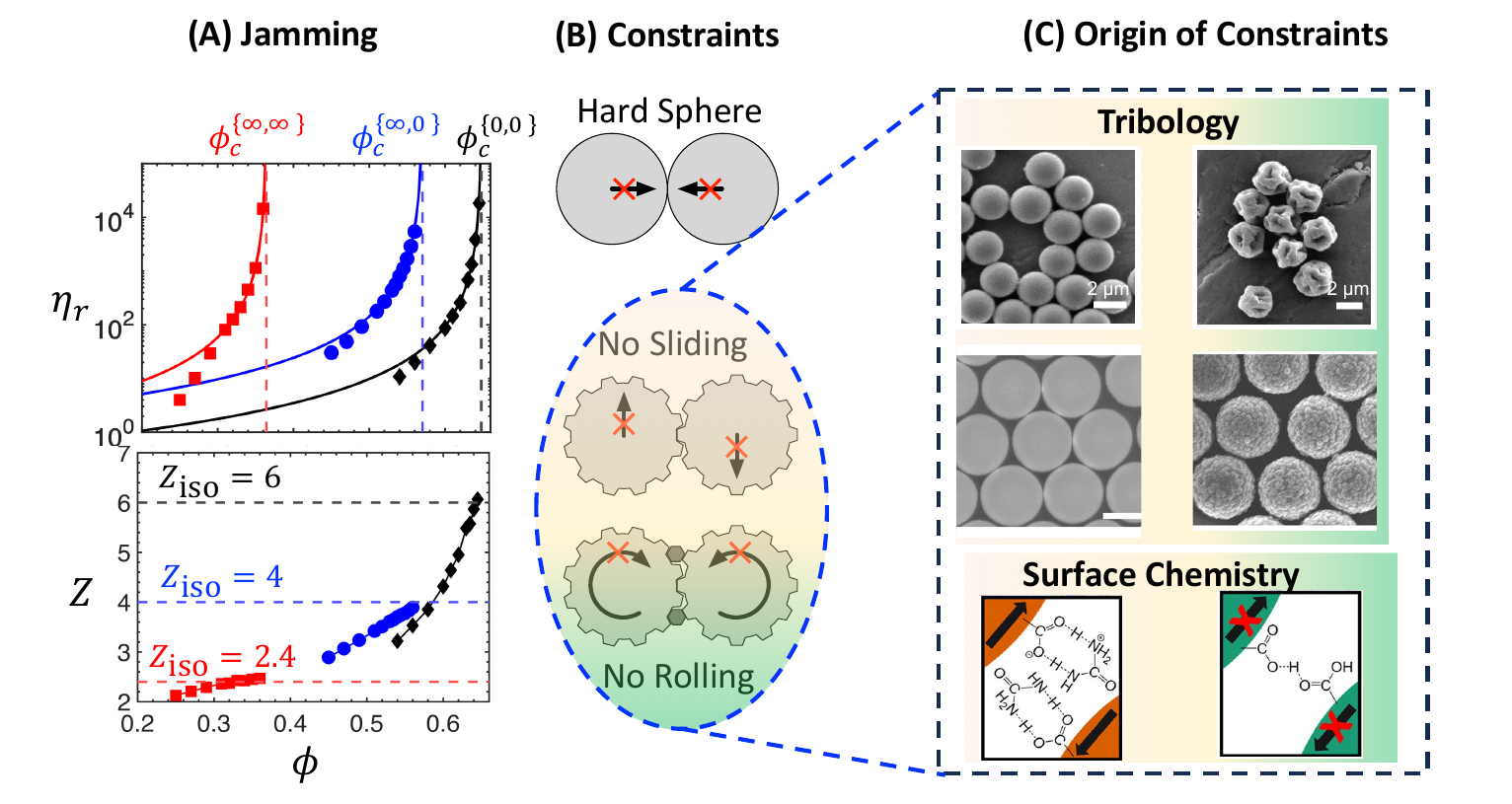}
\caption{\textbf{Relating origin of constraints and rheology.}
(A) Relative viscosity $\eta_r$ ($\eta_r = \eta_s$ the shear viscosity defined in Subsec.~\ref{subsec:suspension_rheology_and_dilatancy}) and coordination number $Z$ vs volume fraction $\phi$. (B) Different types of constraints: hard sphere, $\{\mu_s,\mu_r\} = \{0,0\}$; infinite sliding, $\{\mu_s,\mu_r\} = \{\infty,0\}$; infinite sliding and rolling, $\{\mu_s,\mu_r\} = \{\infty,\infty\}$. (C) Origin of constraints from the particle perspective: constraints can originate from particle surface roughness or surface chemistry, among others.
}
\label{fig:figure_constraints}
\end{center}
\end{figure}

Both static friction and modified hydrodynamics ideas point in the same direction: (i) ideal smooth particles do not exist (at least in the dense limit) and (ii) an additional force, other than hydrodynamics and contact, is needed to observe shear thickening. These two aspects show that shear thickening can be seen as a \emph{stress-activated} crossover from unconstrained to constrained pairwise tangential particle motions, as postulated by Singh \textit{et al.}~\cite{Singh_2020, Singh_2022} and Guy \textit{et al.}~\cite{Guy_2018}. These constraints may arise from many factors: asperity/roughness at the particle level~\cite{Hsu_2018, Hsu_2021}, faceted particle shapes~\cite{Hsiao_2017, Pradeep_2021, dAmbrosio_2023}, interfacial chemistry~\cite{James_2018, James_2019}, among many others~(Fig. \ref{fig:figure_constraints}). In the same spirit as what we discussed in Sec.~\ref{sec:WCmodel}, this crossover happens at an onset stress $\tau_R$. At stresses $\tau \ll \tau_R$, the particles are unconstrained, whereas at stress levels $\tau \gg \tau_R$, the suspension transitions to a fully constrained state, with a higher viscosity.

The original Wyart and Cates model, along with many simulations, are based on static sliding friction with hydrodynamics~\cite{Seto_2013a, Mari_2014, Mari_2015, Mari_2015a, Ness_2016, Boromand_2018, Singh_2018, More_2020, More_2020b}. Singh~\textit{et al.} introduced rolling friction~\cite{Singh_2020, Singh_2022}, thus presenting the following picture. The suspended particles impose (i) hard sphere, (ii) sliding, (iii) rolling, and (iv) twisting constraints, as sketched in Fig.~\ref{fig:figure_constraints}. The hard sphere constraint simply means that no two particles can occupy the same volume. If it is the only one taken into account, any occurring contact is considered to be frictionless. The other terms refer to the constraints on the relative motion between particles, pointing to the theoretical limits of no sliding, no rolling (together with no sliding) and no twisting (with no sliding and rolling). Note that in two-dimensions, only sliding and rolling need to be considered, since the twist mode can only exist in three dimensions. The friction coefficients (or modes) that are introduced for the sliding and rolling constraint terms restrict the relative motion of the particles and stabilize the contact network. Instead of having a single particle--particle friction coefficient, we thus now have both a sliding particle friction coefficient~$\mu_s$ and a rolling particle friction coefficient~$\mu_r$. From this, Maxwell counting arguments~\cite{Hecke_2009,Behringer_2018,Song_2008, Santos_2020} give a lower and an upper bound to the isostatic number of contacts per particle $Z_{\mathrm{iso}}$. For finite values of friction, mechanical stability in dimension $d$ is achieved for $d(d+1)/(2d-1) \leq Z_\text{iso}^{\{\mu_s,\mu_r\}} \leq 2d$. This isostatic condition then modulates the jamming volume fraction $\phi_c^{\{{\mu_s,\mu_r}\}}$, which decreases as the relative motion is constrained. In three dimensions, the jamming volume fraction ranges spans the range $0.365 \leq \phi_c^{\{\mu_s,\mu_r\}} \leq 0.65$ (see Fig.~\ref{fig:figure_constraints}). The upper limit is roughly close to the so-called random close packing (RCP) limit in dry granular physics~\cite{Hecke_2009,Berryman_1983,Visscher_1972}. Simulations conducted in this framework are capable of reproducing CST, DST, and SJ, and shed light on the specific effects of the different types of constraints~\cite{Singh_2020, Singh_2022}.

The premise of this approach is that the physical or chemical origins of the particle-particle interactions are less important than their overall impact in constraining the ability of neighboring particles to move relative to each other. As already mentioned, asperities, faceted particle shapes, adhesive forces etc. all lead to constraints on relative particle motion in sliding, rolling, and twist modes, which in turn lower the jamming volume fraction and enhance shear-thickening behavior. In this spirit, more recently the suspension rheology is beginning to be viewed as a tool to estimate the physics at the particle level~\cite{Singh_2022}. As a consequence, manipulation of, for example, the shape~\cite{dAmbrosio_2023, Hsiao_2017, Pradeep_2021}, deformability~\cite{Chen_2023}, or interfacial chemistry~\cite{James_2018, James_2019} of the particles changes the constraints and affects the bulk rheology, allowing one to tune it. First-order deviations from a spherical shape (without manipulating the chemistry) due to asperities~\cite{Hsu_2018, Hsu_2021}, faceted~\cite{Hsiao_2017, Pradeep_2021} and / or flattened globular faces~\cite{dAmbrosio_2023, Estrada_2008, Estrada_2011}; adhesion due to hydrogen bonding effects~\cite{James_2018, James_2019} or a change in pH~\cite{Laun_1984}, can all be understood in terms of constraints. By design, these concepts cannot however be extended to more complex particle shapes, such as cuboids or super balls~\cite{Cwalina_2017, Cwalina_2016,Royer_2015}.

\section{Beyond a mean-field description: fluctuations and frictional network}
\label{sec:beyond_mean_field}

The mean-field models discussed so far (Wyart and Cates model and constraints-based models) are quite promising, as they reproduce strain-averaged steady-state experimental~\cite{Guy_2015, Guy_2018, Lin_2015, Royer_2016} and simulation~\cite{Singh_2018, Singh_2019, Singh_2020, Singh_2022, Ness_2016, More_2020b, More_2020} data sets for dense shear-thickening suspension flows. However, by construction, they do not consider large-scale or long-range spatial correlations that are often associated with jamming or rigidity transitions in particulate systems~\cite{DeGiuli_2015,Andreotti_2012,Andreotti_2013, Olsson_2007}. They are also agnostic to the geometrical and topological structure of frictional force chains and do not consider the role and emergence of inherent large and intermittent fluctuations in the sample, which we discuss below.

\begin{figure}[!ht]
\begin{center}
\includegraphics[width=\linewidth]{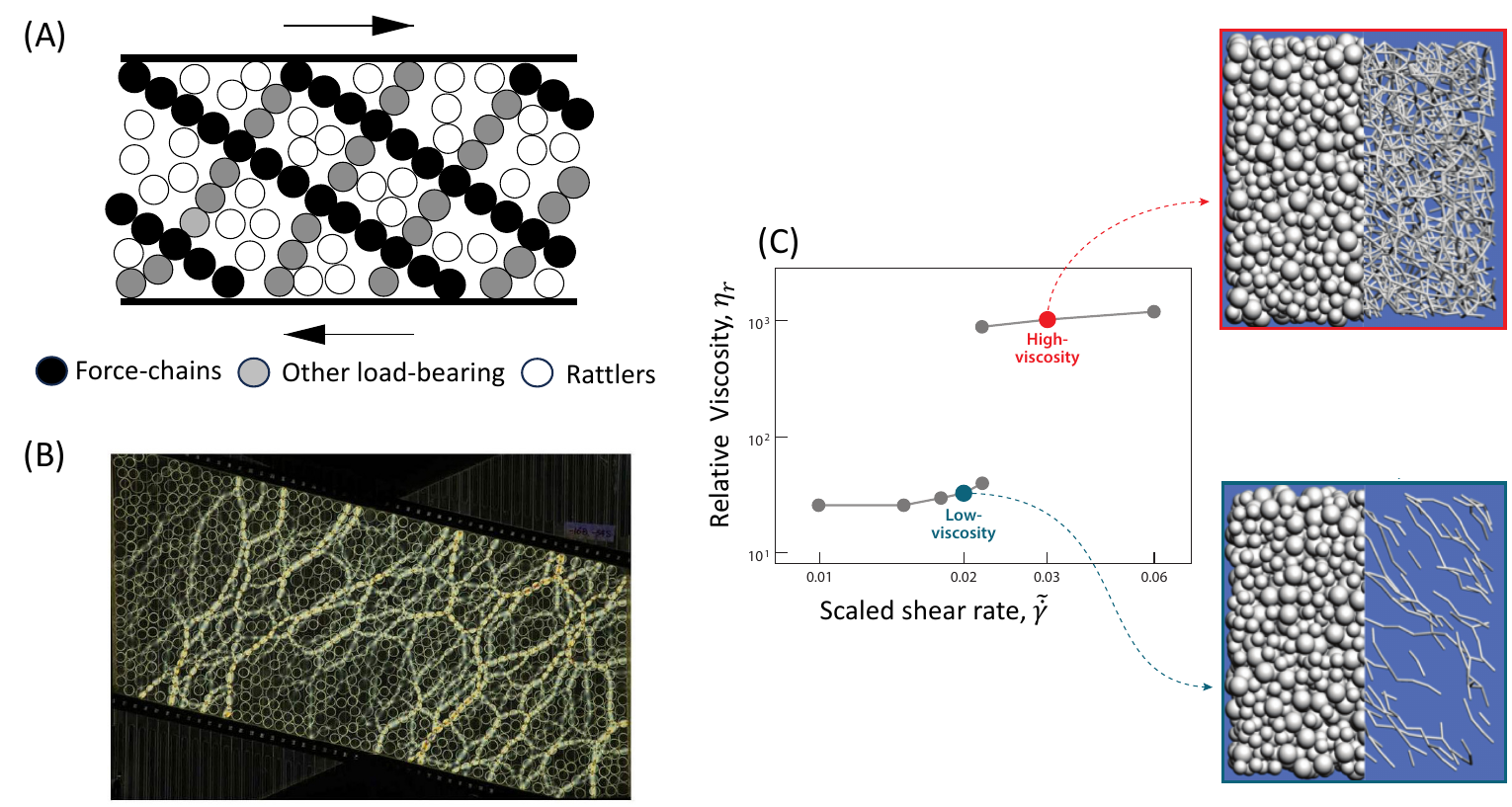}
\caption{\textbf{Jamming, force chains, and discontinuous shear thickening.}
(A) Two-dimensional schematic as proposed by Cates \textit{et al.}~\cite{Cates_1998} showing force chains, other load-bearing and rattler particles, that is, particles that don't participate in frictional contacts. (B) A shear-jammed state in a simple shear experiment using photoelastic discs~\cite{ren2013nonlinear}, where the bright colors show force chains percolating in compressive and tensile directions. (C) Frictional contact network of low-viscosity and high-viscosity states of a shear-thickening suspension; based on simulations from Seto \textit{et al.}~\cite{Seto_2013a}.
}
\label{figure_network_0}
\end{center}
\end{figure}

As discussed in previous sections, shear thickening is driven by stress-activated constraints, which we chose to interpret and model as solid friction. In particulate systems, it is established that friction leads to jamming and shear-induced jamming~\cite{Cates_1998, Bi_2011,Behringer_2018}. Let us therefore start by providing a granular perspective. In granular physics, the emergence of jamming has been shown to be intimately related to the emergence of enduring contacts that lead to robust mesoscale frictional contact and force networks~\cite{Cates_1998, Behringer_2018, Bi_2011, Majmudar_2005}. The concept of force chains dates back to the pioneering work of Cates \textit{et al.}~\cite{Cates_1998}, who defined a force chain as \textit{a linear string of rigid particles in point contact}. They further note that under shear or any external deformation, once the force chains emerge along the principal compressive direction, this structure, together with an orthogonal support from other force-bearing particles, leads the system to jam (or behave like a solid). This is illustrated in Fig.~\ref{figure_network_0} (A). Contemporaneous experiments from Behringer and coworkers experimentally validated these ideas, using photoelastic discs that allowed visualization of the local stress (force) on the particles and popularized the force chain concept. An example of such an experimental system is presented in Fig.~\ref{figure_network_0} (B). Series of studies on photoelastic disks have proven to be useful in obtaining local forces/stresses~\cite{Majmudar_2005}, establishing heterogeneity in particle forces~\cite{Howell_1999}, and experimental validation of shear-jamming~\cite{Bi_2011, Ren_2013,Zhao_2019}.

Previous experimental studies had shown that shear thickening is accompanied by large-scale temporal fluctuations: In the classical rheological perspective, in a rate-controlled set up, shear stress (or viscosity) often fluctuates by orders of magnitude at the critical rate~\cite{Hoffman_1972, Hoffman_1974, Lootens_2003, Lootens_2004, Bender_1996, Boersma_1991, DHaene_1993}. This observation, coupled with the knowledge that frictional contacts are implicated in the emergence of shear thickening, suggests that these large-scale temporal fluctuations should arise from major changes or rearrangements in the force and contact networks~\cite{Lootens_2003, Cates_1998}. This was demonstrated by simulations by Seto and Mari \textit{et al.}~\cite{Seto_2013a, Mari_2014}, who established that a dynamic network of system-spanning frictional contacts is responsible for the high viscosity of the shear-thickened (constrained) state. Furthermore, Mari \textit{et al.}~\cite{Mari_2014} also showed that traditional microstructural responses are subtle across the pure DST regime, that is, from low-viscosity (low stress) to high-viscosity (large stress) flowing states, which does not explain the orders of magnitude change in viscosity. This work paints the following picture for the frictional network (Fig.~\ref{figure_network_0} (C)): At low stress conditions $\tau \ll \tau_R$, all contacts are unconstrained or frictionless. As stress increases, $\tau \approx \tau_R$, frictional contacts form and force chains emerge as roughly linear structures along the compression axis only; this is referred to as low-viscosity condition in Fig.~\ref{figure_network_0} (C). Eventually, at $\tau \gg \tau_R$, the suspension reaches a fully shear-thickened (high-viscosity) state and is found to have force chains along the compressive \emph{and} tensile directions, which stabilize the force chains in the compressive direction and prevent their buckling under shear. These observations directly point to similarity with the propositions of Cates \textit{et al.}~\cite{Cates_1998}, Radjai \textit{et al.}~\cite{Radjai_1998}, and Bi \textit{et al.}~\cite{Bi_2011} and reaffirm the intimate connection between DST and SJ from a frictional network perspective.

\begin{figure}[!ht]
\begin{center}
\includegraphics[width=\linewidth]{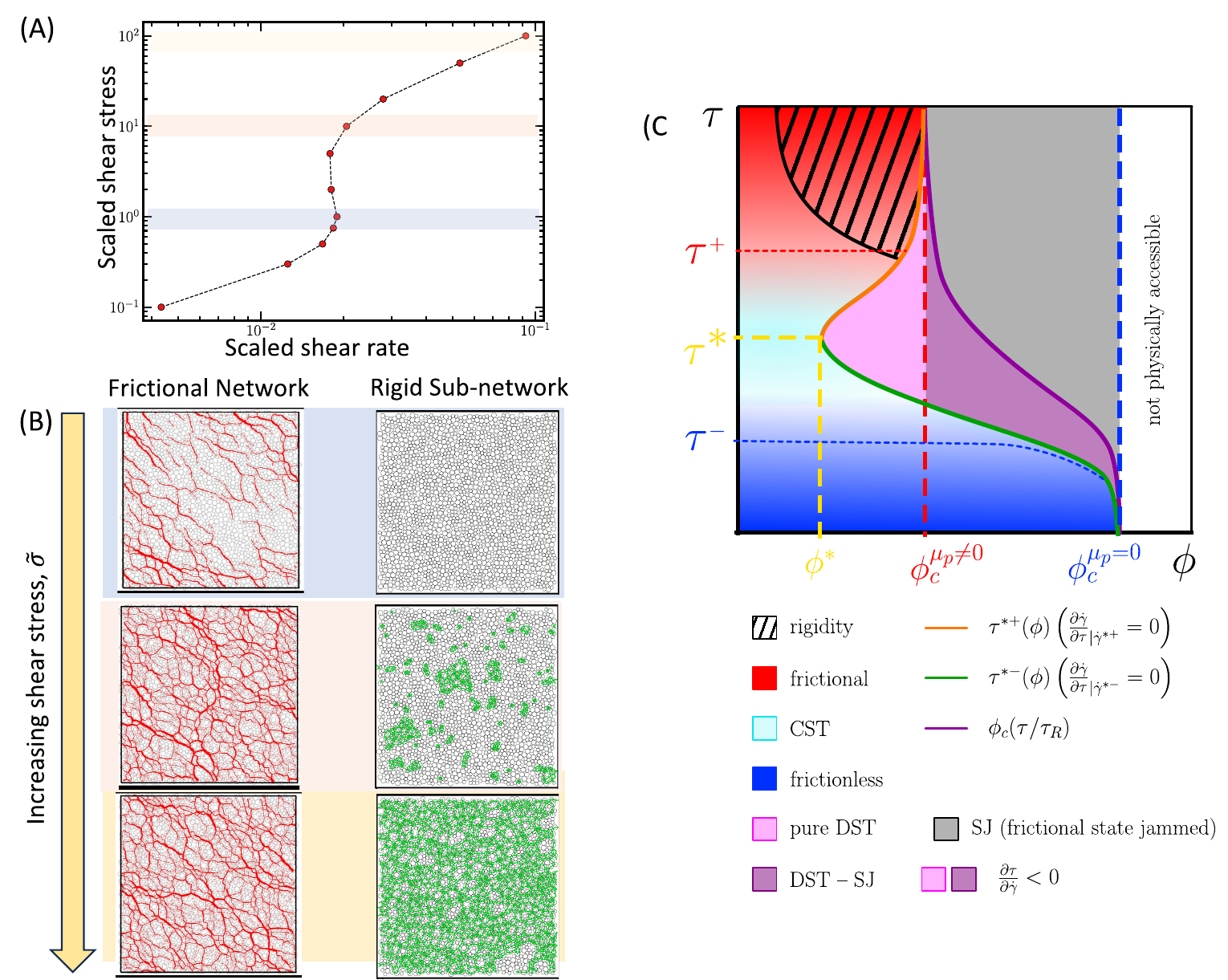}
\caption{\textbf{Frictional contact network, rigidity and SJ.}
(A) $\tau(\dot{\gamma})$ flow-curve showing DST, based on simulations of Naald \textit{et al.}~\cite{Naald_2024}; (B) Frictional network (left) and associated rigid minimally rigid cluster (right) for stress levels that correspond to onset of CST, onset of thickened-state, and fully thickened-state at a packing fraction $\phi>\phi^{*}$, that is, in pure DST regime. (C) Flow state diagram delineating the flow behaviors of shear-thickening dense suspensions in the $\tau$--$\phi$ space. The lines separate different flow regimes based on the mean-field Wyart-Cates model: shear-jammed (grey), pure DST (pink), DST-SJ (purple), and isotropic frictionless jamming (black). DST and SJ lines are based on the Wyart-Cates model giving the locus of, respectively, states with $\frac{\partial \dot{\gamma}}{\partial \tau} = 0$ and $\phi_c(\tau/\tau_R)$ (Eq.~\eqref{eq:phi_c_Pi}). Blue and red colors show frictionless and fully frictional states. The dashed region with black lines indicate the system-spanning minimally rigid states. Adapted from Naald \textit{et al.}~\cite{Naald_2024}.}
\label{figure_network_1}
\end{center}
\end{figure}

Thomas \textit{et al.}~\cite{Thomas_2018, Thomas_2020} showed that shear thickening is accompanied by the emergence of correlations in the force space. As previously stated, Mari \textit{et al.}~\cite{Mari_2014} demonstrated that traditional correlations show only subtle differences across pure DST. This means that the network space of frictional contacts and forces, rather than the physical space of particles, holds the key to providing a mechanistic understanding of shear thickening. This led to studies using network science and graph theory tools to analyze the topology and geometry of frictional contact and force networks~\cite{Sedes_2020, Sedes_2022,Gameiro_2020,Boromand_2018,Naald_2024,dAmico_2025,Sharma_2025, Edens_2019, Edens_2021, Boromand_2018, Nabizadeh_2022}. In the context of network science, the physical space of particles is converted into a graph, where each particle is represented as a node, and the frictional contact between particles is represented by an edge. Gameiro \textit{et al.}~\cite{Gameiro_2020} demonstrated that pure DST is accompanied by the emergence of loop-like structures in the frictional network. The authors used persistent homology to describe the topology of the frictional force network and used total persistence as a measure of the importance of loops (see below) to link the force network with the rheology. The correlation was found to be independent of volume fraction $\phi$ and applied stress $\tau$, suggesting that viscosity is solely controlled by the persistence of the frictional network. In the same spirit, D'Amico \textit{et al.}~\cite{dAmico_2025} decomposed the frictional contact network into loops. In network science, a loop (or cycle) is defined as a closed path in a graph, which starts and ends at the same node, traversing a sequence of edges and nodes without repeating any edge or node (except the starting/ending node)~\cite{Papadopoulos_2018}. Thus, a loop of order $l$ can be thought of as a polygon with $l$ sides. They showed that the suspension's relative viscosity $\eta_r$ can be expressed in terms of the number of third-order loops $n_3$. Moreover, this correlation is insensitive to the full phase space $(\tau,\phi,\mu_s)$ that controls the rheology of the dense suspension. This has deep connections with the rigidity concepts in graph theory. According to Laman's theorem~\cite{Laman_1970}, triangles are found to be the smallest rigid structures, that is, they do not deform under an externally applied load. More recent work has attempted to probe the concepts of rigidity of the local frictional contact network and relate it to shear thickening. Using different methods, Goyal \textit{et al.}~\cite{Goyal_2024} and Naald \textit{et al.}~\cite{Naald_2024} probed the local rigidity of the frictional network and found the onset of system-spanning minimally rigid clusters in the vicinity of $\phi^{*}$, the volume fraction that marks the onset of DST. The frictional contact network and the associated minimally rigid clusters for different stress levels $\tau/\tau_R$, at a packing fraction $\phi>\phi^{*}$, that is, in the pure DST regime, are illustrated in Fig.~\ref{figure_network_1} (B). In particular, Naald \textit{et al.}~\cite{Naald_2024} suggested that the onset of rigidity drives the enhanced dissipation allowing for high viscosities $\eta_r > 1000$, thus showing that rigidity (rather than DST) is a precursor to SJ.

Sedes \textit{et al.}~\cite{Sedes_2022} used the K-core network to analyze the contact network clusters in a percolated network, as the suspension shear thickens. The $k-$cores are subgraphs (or ``clusters'') of the frictional subnetwork, where all nodes have $k$ or more contacts with other nodes in the subgraph. Using the Molloy-Reed criterion~\cite{Molloy_1995}, the authors showed that percolation occurs in both CST and DST, which means that percolation is not the discriminating measure for strong CST and DST. They defined $\chi_\sigma = \frac{d\sigma}{d\dot{\gamma}}$, a rheological susceptibility, and showed that the appearance of $3$-core structures coincides with the divergence of $\chi_\sigma$. This hints at the onset of DST (that is, the $(\tau^{*}, \phi^{*})$ point in the state diagram) being a non-equilibrium critical point, an aspect which needs to be further pursued. In this direction, more recently Ramaswamy \textit{et al.}~\cite{Ramaswamy_2023} have noted that shear thickening behavior is instead governed by two separate critical points, frictional and frictionless jamming points; demonstrating
a crossover scaling framework to characterize shear thickening in cornstarch and silica suspensions. Finally, Nabizadeh \textit{et al.}~\cite{Nabizadeh_2022} used community detection to analyze the frictional contact network, and found that the force clusters formed in the pure DST regime are more tightly constrained by their surrounding clusters than the ones in the CST regime.

Results from network science allow to populate the state diagram from Fig.~\ref{fig:WC_Model} (as originally proposed by Wyart and Cates), presented here in Fig.~\ref{figure_network_1} (C). In this form of diagram, a new rigidity line emerges where minimally rigid clusters begin to span the system. This state in shear-thickening suspensions lies below the onset of SJ and we believe is the true precursor to SJ, rather than DST, as previously hypothesized~\cite{Peters_2016, Morris_2020}.

To conclude this section, let us mention that the effect of constraints (sliding and rolling modes) on the rheology is also beginning to be established. How it affects the frictional network has, however, not yet been explored. Preliminary work suggests that systems with similar viscosity originating from various combinations of $\{\mu_s,\mu_r\}$ can have very different frictional networks~\cite{Sharma_2025}. However, the proposition of Cates \textit{et al.}~\cite{Cates_1998} (Fig.~\ref{figure_network_0} (A)) only considers sliding constraints and thus an orthogonal support is needed for the force chains to not buckle under external deformation. More work is needed to consider how constraints affect the minimal requirement for the force chains or the particulate system to be mechanically stable.

\section{Beyond simple shear: realistic flows and fluid dynamics}
\label{sec:funky_experiments}

In the previous sections, we discussed mean-field models such as the Wyart and Cates model, as well as some of their limitations, and pointed to network science tools and how they allow us to build our understanding of dense suspensions rheology when mean-field models are no longer enough. In the context of steady-state flows and either in simple shear or in rheometer experiments designed to be as close as possible to a simple shear flow, mean-field models have been shown to capture both experimental~\cite{Guy_2015, Guy_2018} and simulation data sets~\cite{Mari_2014, Ness_2016, Singh_2018, Singh_2020}. Additionally, even in an ``ideal'' flow in a rheometer, the presence of S-shaped flow curves or the onset of DST leads to instabilities, shear banding, and spatiotemporal fluctuations~\cite{Hermes_2016, Saint-Michel_2018}. In this section, we discuss recent experimental studies that explore the behavior of shear-thickening suspensions in even more complex types of flow. In particular, we will discuss the extent to which the Wyart and Cates model can still provide some basis to the analysis of these flows.

\subsection{Flow down an incline}
\label{subsec:flow_down_incline}

Flow down an incline is among the archetypical classic problems in fluid mechanics, where the Reynolds number framework based on pure fluid predicts Kapitza~\cite{Kapitza_1948} or roll-waves instabilities due to turbulence (inertia) in open channels. Such instabilities are also observed in other types of complex fluids, such as power-law fluids~\cite{Hwang_1994, Allouche_2017}, mud flow, and granular materials~\cite{Forterre_2003, Forterre_2006}. In all of these cases, the instability is driven by inertia.

\begin{figure}[!ht]
\begin{center}
\includegraphics[width=\linewidth]{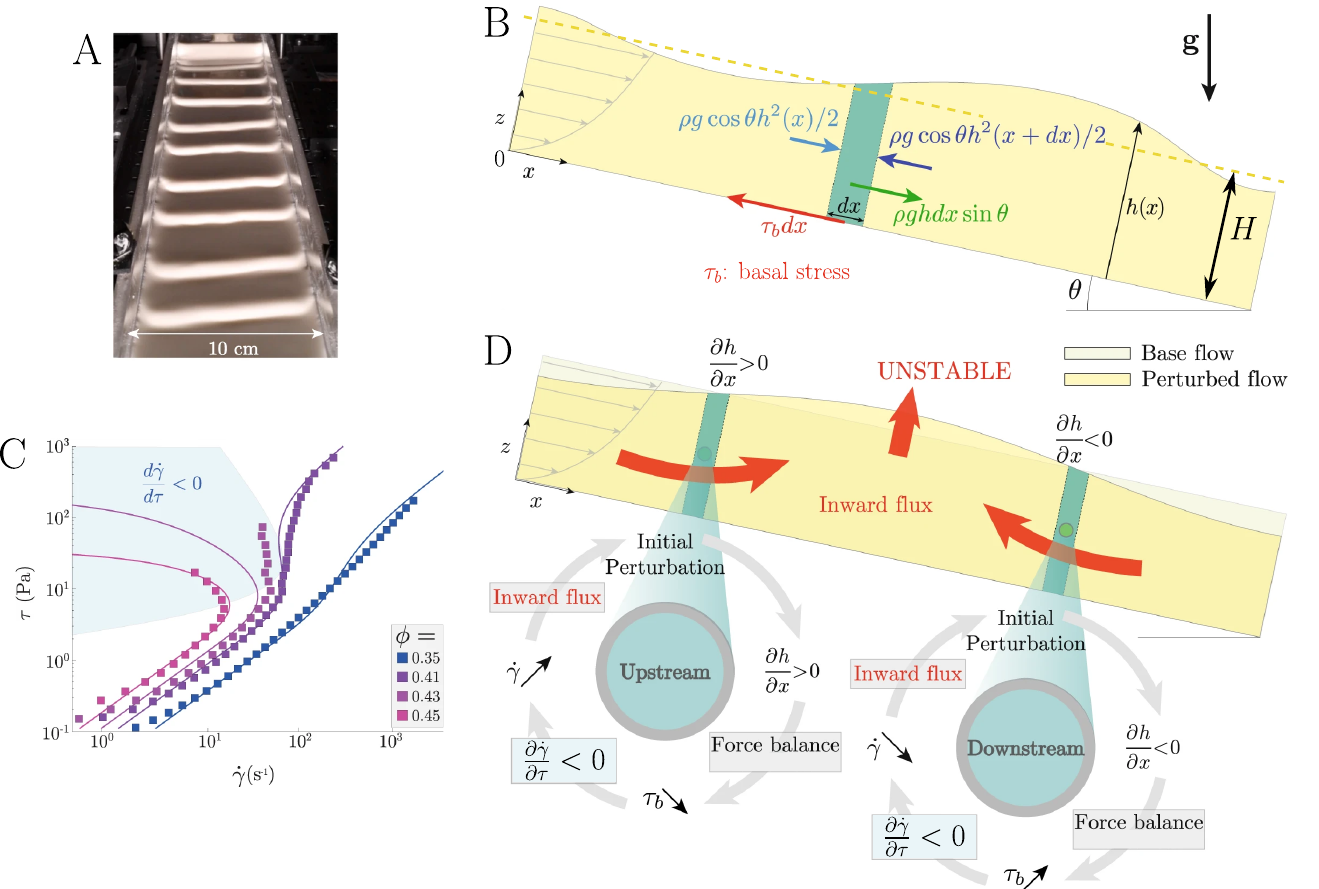}
\caption{``Oobleck wave'' down an incline plane. A: Photo of the developping waves. B: Sketch of the problem. C. Measured rheology, and fits to the Wyart and Cates model. D. Oobleck waves mechanism. Adapted from~\cite{DarboisTexier_2020} (article and figures licensed under a Creative Commons Attribution 4.0 International License, copy of the license: https://creativecommons.org/licenses/by/4.0/).}
\label{fig:flow_down_incline}
\end{center}
\end{figure}

In the case of a Newtonian liquid, the force of gravity down the inclined plane can be written as $g \sin(\theta)$, where $\theta$ and $g$ are the angle of plane and acceleration due to gravity, respectively. This would be balanced by the viscous force $\eta u_{zz}$, where $\eta$ is the fluid's viscosity. Here, we focus on a two-dimesional Cartesian coordinate system with $z=0$ denoting the inclined plane and $x$ pointing downslope, with $(u,w)$ the velocity field (see Fig.~\ref{fig:flow_down_incline} B.). This implies $\eta U/\rho H^2 \sim g\sin(\theta)$, that is, $\Reynolds = \rho UH/\eta \sim U^2/(gH\sin(\theta))$, with $\rho$ the fluid's density. In this case, the Kapitza threshold is expected to be $\Reynolds_K = 5/(6\tan(\theta))$, which is found to be such that $\Reynolds_K \gg 1$ for the typical tilt angles studied.

Earlier experimental observations of the flow of a shear-thickening suspension down an incline by Balmforth \textit{et al.}~\cite{Balmforth_2005} showed the occurrence of instability waves at Reynolds number $\Reynolds \lessapprox 1$, much smaller than expected and therefore unexplained. The authors concluded their article by ``\textit{Whatever the physical origin of the observed phenomenon, we hope that our study will motivate further investigations of this curious material.}''. This problem has recently been revisited by Texier \textit{et al.}~\cite{DarboisTexier_2020, DarboisTexier_2023} in the context of material instability that originates from the rheology of cornstarch and its coupling with the free surface. They reported the emergence of an unstable kinematic wave termed the ``Oobleck wave'' that originated at the onset of DST $(\tau^{*},\phi^{*})$, as predicted by the Wyart and Cates model. The S-shaped flow curves with negative slope (see Fig.~\ref{fig:flow_down_incline} C.), coupled with the free surface, trigger an instability in the medium that does not require inertia. A mismatch between hydrostatic pressure and basal stress $\tau_b$ leads to the amplification of the kinematic surface wave, and the unstable mode propagates at the speed of the surface kinematic waves. The mechanism proposed in~\cite{DarboisTexier_2020} is illustrated in Fig.~\ref{fig:flow_down_incline}.

\subsection{Flow through a constriction and liquid migration}
\label{subsec:flow_through_constriction}

\begin{figure}[!ht]
\begin{center}
\includegraphics[width=\linewidth]{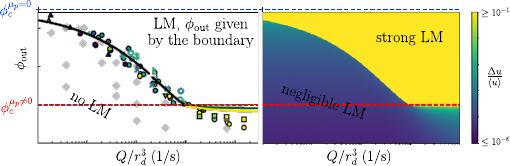}
\caption{Liquid migration in the flow of a shear-thickening suspension through a constriction. Left: experimental data. Right: one-dimensional model, using interpolated jamming packing fraction from the Wyart and Cates model. The quantity $\Delta\langle u \rangle/u$ quantifies the particle phase motion relative to the suspending fluid. Adapted from~\cite{O'Neill_2019}.}
\label{fig:flow_constriction}
\end{center}
\end{figure}

Flow of dense particulate materials through an extruder or through a constriction often leads to liquid migration (LM)~\cite{O'Neill_2019} (also known as self-filtration~\cite{Haw_2004}). Solids that build up near the constriction can cause jamming, thus impeding the flow, and the material becomes more dilute. O'Neill \textit{et al.}~\cite{O'Neill_2019} performed extrusion experiments on a mixture of cornstarch and water, and coupled their findings with stress-controlled steady-shear rheology. They showed that the volume fraction $\phi_{\mathrm{out}}$ of the extruded material is simply controlled by the scaling variable $Q/R_d^3$, with $Q$ and $R_d$ being the volumetric flow rate and the radius of the die, respectively. The term $Q/R_d^3$ sets the shear rate scale in the die. The LM state diagram shows the master curve as a ``phase boundary'' for different initial volume fraction $\phi_{\mathrm{in}}$, barrel radius $R_b$ and die radius $R_d$ (see Fig.~\ref{fig:flow_constriction}). For low volume fractions $\phi_{\mathrm{in}}$ of input material, $\phi_{\mathrm{out}} \simeq \phi_{\mathrm{in}}$. For an initial volume fraction $\phi_{\mathrm{in}}$ larger than a certain critical value $\phi_{\mathrm{in}}^c$, liquid migration occurs: $\phi_{\mathrm{out}} < \phi_{\mathrm{in}}$. Using steady-shear rheology, $\phi_{\mathrm{in}}^c$ was found to coincide with the onset of DST $\phi^{*}$ as predicted by the Wyart and Cates model.

\subsection{Flow through a pipe}
\label{subsec:flow_pipe_frictional_soliton}

\begin{figure}[!ht]
\begin{center}
\includegraphics[width=\linewidth]{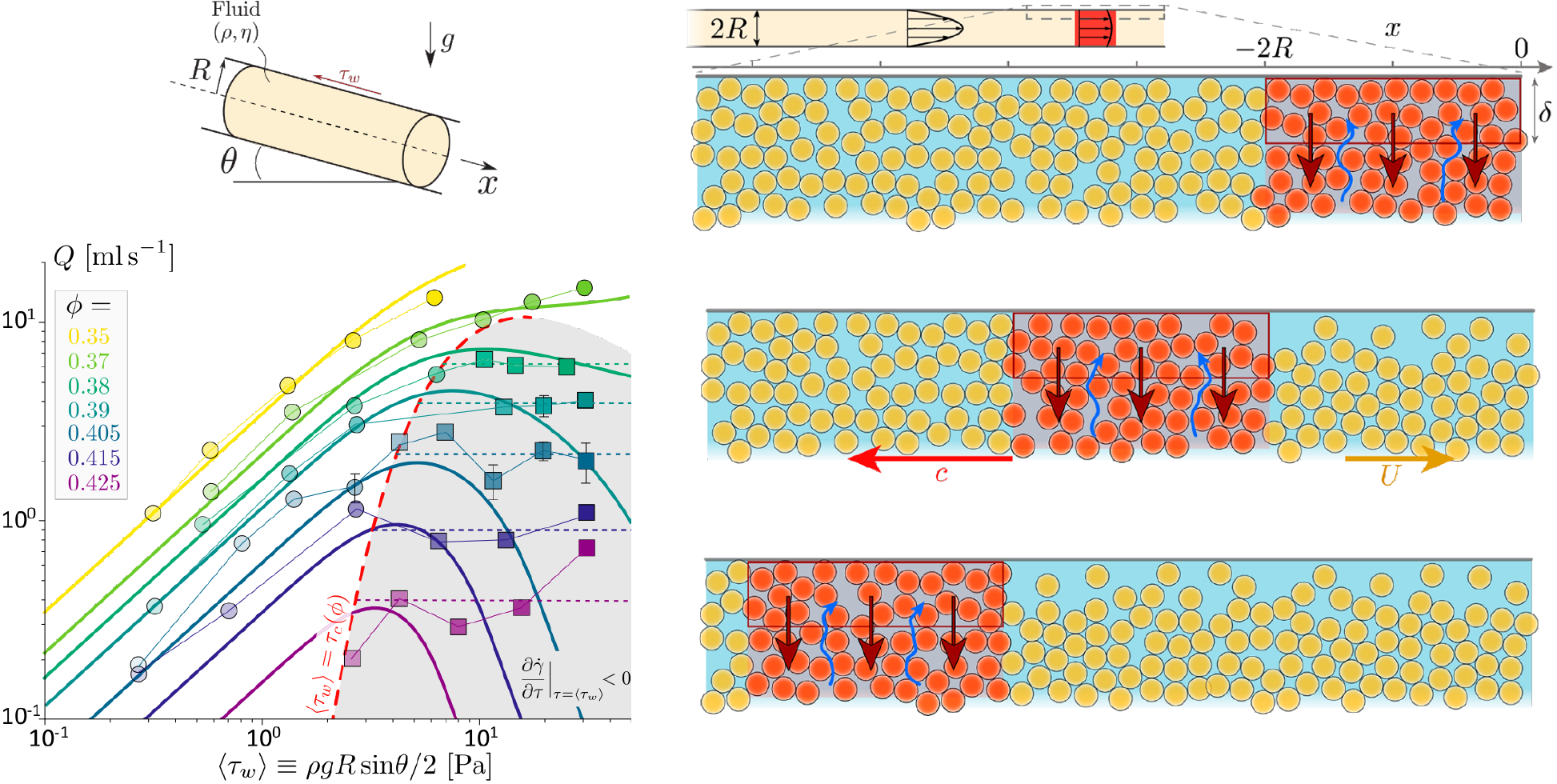}
\caption{Flow of shear-thickening suspensions through a pipe: flow-rate saturation due to the nucleation and propagation of a frictional soliton. Top left: Sketch of the problem. Bottom left: flow rate saturation, comparison between the experimental data (symbols) and the predictions of the Wyart and Cates model (lines). Right: Proposed mechanism for the nucleation and upstream propagation of the frictional soliton. Adapted from~\cite{Bougouin_2024} (article and figures distributed under Creative Commons Attribution License 4.0 (CC BY)).}
\label{fig:soliton}
\end{center}
\end{figure}

Pipe flow is one of the most usual chemical engineering problems, often encountered in industrial and natural settings. In the case of a Newtonian liquid of viscosity $\eta$ flowing through a pipe of radius $R$ with a pressure gradient $\Delta P$, the flow rate in the laminar regime is $Q \propto R^4\Delta P/\eta$. Bougouin \textit{et al.}~\cite{Bougouin_2024} studied the case of shear-thickening suspensions, where they observed some intriguing physics. They showed that for volume fractions $\phi < \phi^{*}$, the flow rate increases linearly with the mean shear stress at the wall, as expected from a laminar flow. Beyond the onset of DST, at $\phi>\phi^{*}$, and at wall stress $\tau_w$ levels below $\tau^{*-}(\phi)$, the flow rate still increases linearly, like that of a Newtonian liquid. The near-wall flow is uniform and steady and almost an order of magnitude smaller than the mean-flow velocity, in line with what is observed for a Poiseuille-like laminar flow. At stress levels $\tau_w > \tau^{*-}(\phi)$, the flow rate becomes independent of forcing (even for an increase of an order of magnitude). This is illustrated in Fig.~\ref{fig:soliton}, bottom left. A spatiotemporal inspection reveals the propagation of a frictional soliton that separates two laminar flow regions (illustrated in Fig.~\ref{fig:soliton}, right). In these regions, the velocities are orders of magnitude below the mean velocity, as in the low-forcing regime. The frictional soliton however is a plug-like flow. The authors show that the frictional soliton is longitudinally localized, propagates upstream, and preserves its shape spanning the cross sections of the pipe, with the dissipation concentrating over a flow length of the order of pipe radius. They propose a minimal mechanism for the upward propagation of the soliton that is based on a transient Reynolds-like dilation and a subsequent Darcy backflow. This mechanism is illustrated in Fig.~\ref{fig:soliton}, left. The coupling between, on one hand, transient dilation in a shear-thickening suspension forced into its frictional state, and on the other hand, the subsequent Darcy backflow is further discussed in~\cite{Athani_2025}, showing that while frictional contacts dominate the steady-state behavior of shear-thickening suspensions, the fluid phase and hydrodynamics are very important to describe transients.

Once again, the Wyart and Cates model predicts the onset of the deviation from the typical Newtonian behaviour. However, it incorrectly describes what happens once the flow rate saturates. This is to be expected, since in this regime the suspension is separated into two distinct phases, while the Wyart and Cates model considers the suspension as one continuous phase.

\subsection{Air injection in a shear-thickening suspension}
\label{subsec:bubbles_and_air_invasion}

\begin{figure}[!ht]
\begin{center}
\includegraphics[width=\linewidth]{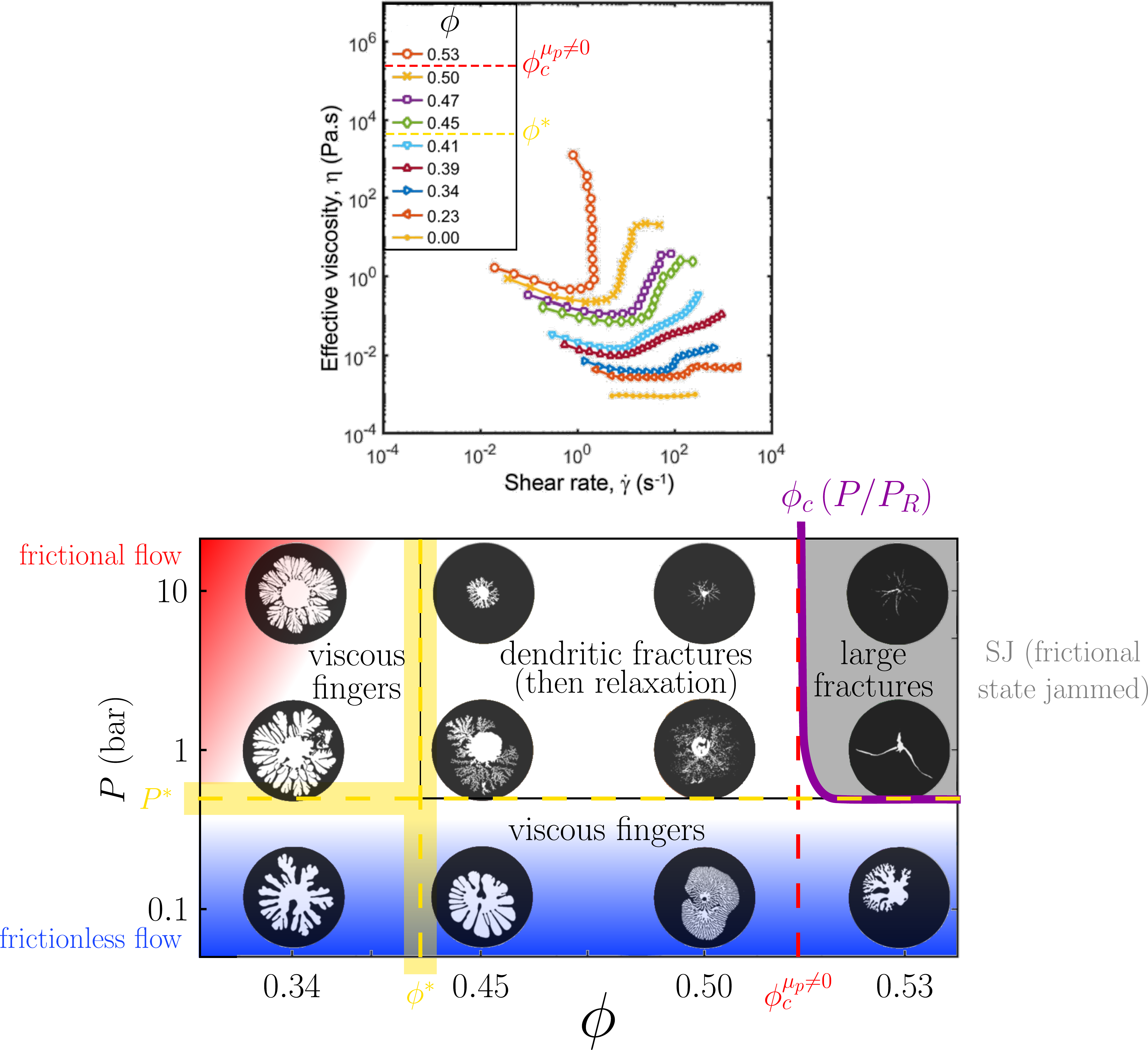}
\caption{Air invasion in a Hele-Shaw cell filled with suspensions of cornstarch in water. Top: measured rheological curves. Bottom: phase diagram of the different response regimes, in the injected pressure $P$-packing fraction $\phi$ plane. The pure DST, DST-SJ, and rigidity regimes cannot be placed here, and the positions of $P^{*}$ (equivalent of $\tau^{*}$) and $\phi^{*}$ are only estimated. The purple line is the equivalent of that of Fig.~\ref{fig:WC_Model}, with $P_R$ being the equivalent of $\tau_R$. Adapted from~\cite{Ozturk_2020} (article and figures licensed under a Creative Commons Attribution 4.0 International License, copy of the license: https://creativecommons.org/licenses/by/4.0/).}
\label{fig:hele_shaw}
\end{center}
\end{figure}

Two different papers, by Ozturk \textit{et al.}~\cite{Ozturk_2020} and Lilin \textit{et al.}~\cite{Lilin_2023}, explored the effect of air injection into a shear thickening suspension. In~\cite{Ozturk_2020}, the authors study the problem of air invasion in a Hele-Shaw cell (see Fig.~\ref{fig:hele_shaw}). They first measure the steady-state rheological flow-curve of cornstarch suspensions in water, and use this to estimate the values of $\phi^{*}$ and $\phi_c^{\mu_p\neq 0}$, the packing fractions marking the onset of the DST and SJ regimes respectively (see Fig.~\ref{fig:hele_shaw}, top). They show that below a critical air injection pressure $P^{*}$ (equivalent of $\tau^{*}$ as defined in Sec.~\ref{sec:WCmodel}), the suspension exhibits viscous fingering, as would be the case for a Newtonian fluid. Above $P^{*}$, the behavior depends on the suspension's packing fraction. Below $\phi^{*}$, it still exhibits viscous fingering, behaving like a Newtonian fluid. Between $\phi^{*}$ and $\phi_c^{\mu_p\neq 0}$, the authors observe what they call dendritic fracturing. Upon air invasion, the system fractures locally into a dendritic network, which then relaxes partially. The pattern is conserved, but the edges of the initial fractures are smoothed out. This can be rationalized using the Wyart and Cates model. Upon air invasion, if the air pressure is larger than $P^{*}$, the system is locally forced into its frictional state, and transiently jams, as would a suspension of frictional hard spheres subjected to a sudden high stress. Once the high stress front has passed, the suspension can relax to a liquid state, because its frictional, high viscosity branch still has flowable states. As for the frictional soliton discussed in Subsec.~\ref{subsec:flow_pipe_frictional_soliton}, the transient behavior here is quite important to understand the entire response of the suspension, and ideas described and developed in~\cite{Athani_2025}, particularly the effect of particle confinement, allow to rationalize the observations. Then, above $\phi_c^{\mu_p\neq 0}$, the authors observe a large-scale fracturing regime, in which the system also fractures upon air invasion, but forms a fracturing pattern with a much lower number of branches, that doesn't relax. At these packing fractions, the air pressure prevents the system from evolving on its frictionless branch, so it is forced to stay in a shear-jammed frictional state.

In~\cite{Lilin_2023}, the authors study the formation, growth, and rise of bubbles in a dense cornstarch in water suspension. They show that at low enough mass fractions, the bubble behavior is the same as in a Newtonian liquid, while at very large mass fractions the suspension always locally jams and fractures upon air injection. In this case, the ``bubble'' is really a growing fracture. At intermediate volume fractions, they observe an interesting behavior: at short times, the local shear rate at the bubble border is larger than the critical shear rate $\dot{\gamma}$ above which the suspension shear-thickens, which leads to the suspension initially fracturing. At larger times, the local shear rate $\dot{\gamma}_\text{loc}$ at the bubble border decreases. Indeed, $\dot{\gamma}_\text{loc} \propto \dot{\Omega}/\Omega$, with $\Omega$ being the bubble volume. As time passes, the bubble volume increases, and the authors show that $\dot{\Omega}$ is time-independent: $\dot{\gamma}_\text{loc}$ thus necessarily decreases with time. If this decrease is large enough, the suspension at the border of the bubble relaxes, leading to the rise of a bubble which was initially a fracture. Again, this can be rationalized within the Wyart and Cates model, in the same way as for the previous paper.

\begin{figure}[!ht]
\begin{center}
\includegraphics[width=0.95\linewidth]{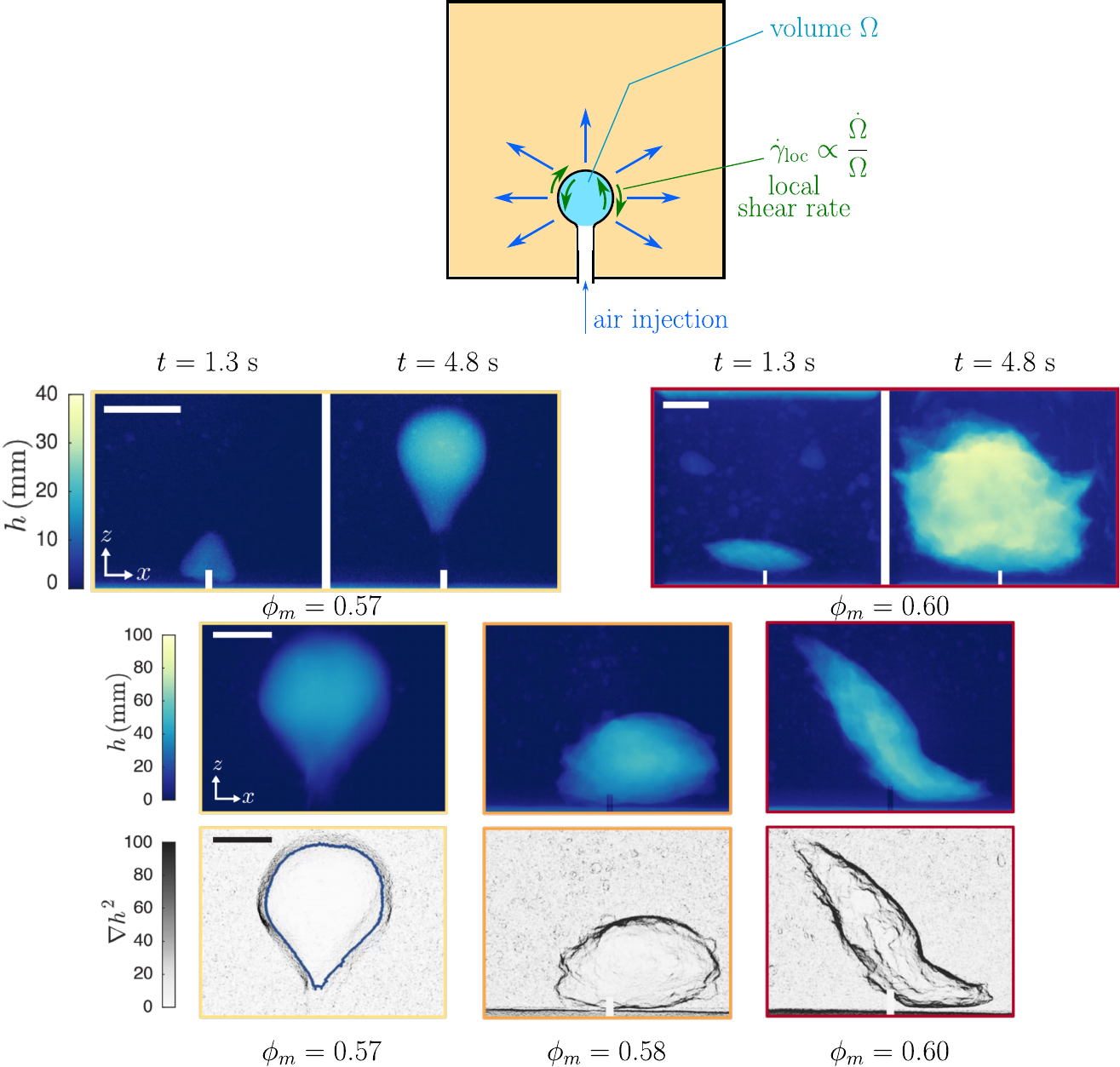}
\caption{Bubble formation, growth, and in certain cases rise, in suspensions of cornstarch in water, for different mass packing fractions $\phi_m$.Top: Sketch of the experimental system. Middle: two different types of bubbles, depending on the mass packing fraction, and their evolution in time. Bottom: comparison of different bubbles and their surface roughness (quantified by $\nabla h^2$), depending on the mass packing fraction, at a fixed time. Adapted from~\cite{Lilin_2023} (article and figures distributed under the terms of the Creative Commons CC BY).}
\label{figure_bubbles}
\end{center}
\end{figure}

\section{Conclusion and perspectives}

In this perspective article, we have discussed some of the recent advances and current understanding of shear thickening (and particularly DST) observed in dense suspensions. The first studies date back to the 1930s, and it is surprising that the mechanistic basis still remains debated. Still, a minimal model picture has emerged to explain this behavior: a two-state model involving a stress-activated transition, at a certain stress-scale $\tau_R$ resulting from the presence of a short-range repulsion between the particles, from unconstrained to constrained pairwise relative tangential motion. This model can \emph{quantitatively} predict shear thickening. This general picture points to distinct physics driving the rheology at low ($\tau \ll \tau_R$) and high ($\tau \gg \tau_R$) stresses, i.e., unthickened and thickened states. The low-stress ``unconstrained'' state is that of classical hydrodynamics with the lubrication film between the particles being intact, while in the high-stress state the lubrication breaks down and the particles come close enough to constrain the pairwise relative motion.

\begin{figure}[!ht]
\begin{center}
\includegraphics[width=0.9\linewidth]{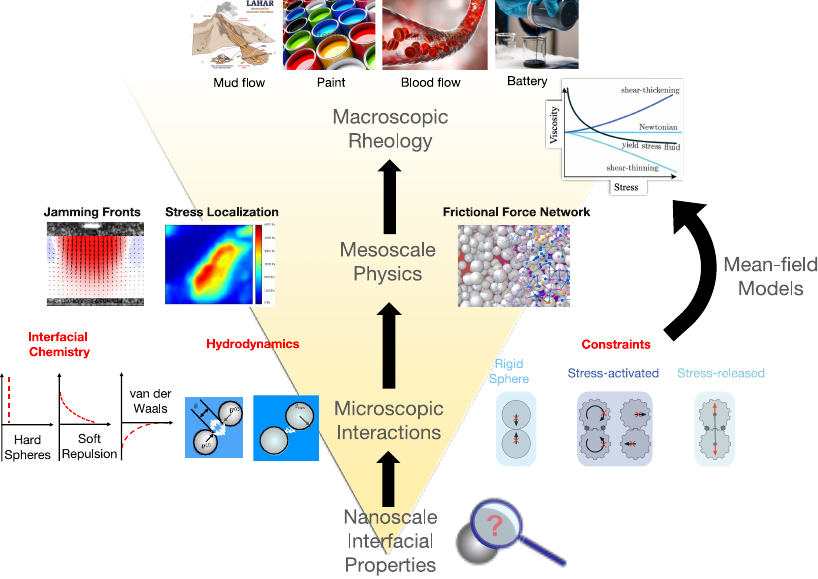}
\end{center}
\caption{\textbf{A hierarchical multiscale view of dense suspensions}. Dense suspensions rheology is measured at the macroscopic scale (steady-state strain averaged observation), and can be understood in terms of a hierarchical interrelationship across length-and time-scales. At the smallest length scale, we have a single particle and surrounding fluid, where asperities or roughness (orders of magnitude smaller than the particles) and interfacial chemistry (with the fluid) lead to microscopic interactions. These microscopic interactions are present at the level of two or more particles and lead to the collective motion of particles, which often originate from constraints between particles. Finally, the correlated motion or the collective behavior (order of $10$ or $20$ particles) dictates the behavior of dense suspensions.
}
\label{Hierarchy_Fig}
\end{figure}

For dense suspensions in general, this paints a picture that we illustrate in Fig.~\ref{Hierarchy_Fig}. At the smallest scale of the problem is a single particle immersed in a fluid, where roughness, faceted shapes, surface irregularities, adhesion \dots~ in summary: surface properties, either intrinsic to the particle or rising from particle--fluid interactions, constrain the relative motion between two particles. These microscopic constraints lead, at high enough volume fraction, to correlated non-affine motion at the mesoscale, which can be quantified in terms of frictional force and contact networks. The behavior of these networks, which are as stated mesoscale objects, in turn determines the macroscopic rheology of the suspension. In the case of shear-thickening suspensions, the constraints only play a role at high enough stresses, that is, for $\tau \gg \tau_R$.

The state-of-the-art constitutive models for shear-thickening suspensions are mean-field in nature, which means they relate the microscopic constraints directly to the macroscopic bulk response of the material. This inherently assumes that the mean simple shear flow exists; thus ignoring many effects: fluctuations, instability, wall slip, free surface, and presence of boundary. The modeling approach, even though mean-field in nature and devoid of any boundary effects, still interestingly predicts the onset of material instability in various non-viscometric flows, as briefed in Section~\ref{sec:funky_experiments}. Of course, it cannot (and should not be expected to) predict the local material state, which would be highly heterogeneous in such cases.

The ideas developed so far only hold for spherical quasi-monodisperse particles (single peaked size), which do not exist in reality. The Wyart and Cates formalism in its original form breaks down with a modest polydispersity in size, although some progress has been made to expand it~\cite{Singh_2024, Guy_2020, Pednekar_2018, Malbranche_2023}. Interestingly, suspensions of quasimonodisperse particle sizes considering spherical (\cite{Guy_2015, Guy_2018, Royer_2016}) or irregularly shaped (cornstarch or mineral crystals~\cite{Richards_2021}) particles all show qualitatively similar shear-thickening behavior. However, the theoretical and simulation understanding in terms of jamming and isostaticity is limited to spherical shape. Simulation and theoretical studies that go beyond spherical shapes will be particularly useful, but challenges associated with both implementation of non-spherical granular particles and the correct lubrication forces associated with face-face, edge-edge, and face-edge interactions impede the development.

Real-world application to multiphase flows will also involve understanding the effect of boundaries or walls, which are present in extrusion of dense paste or redox flow batteries, as well as open-surface (e.g. flow down incline) boundary conditions. In the dense limit, wall slip, localization near walls, particle migration, etc. will be important -- none of which is currently considered theoretically. Additionally, current modeling approaches do not consider transients. However, in real-world flows these need to be considered~\cite{Bougouin_2024, Balmforth_2005,Athani_2022,Athani_2025}. Another point to consider is that in order to apply these concepts to multiphase flows, additional effects like surface tension, capillary effects, etc. need to be considered as well.

Finally, dense suspensions also present a model system to probe the statistical framework for out-of-equilibrium driven systems, which remains an open question. Some recent studies suggest that DST (in a rate-controlled framework) is a critical point. Extending ideas related to criticality and floppy modes from glassy and dry granular physics to dense suspensions can open exciting avenues, as the last decade has witnessed. In this spirit, generalizing the isostatic conditions to a broad polydisperse setting will be extremely useful.

\section{Take-home messages}

In this final section, we list some of the important take-home messages we wish to put forward in light of everything discussed so far in this perspective.

\begin{enumerate}

    \item Now that things are becoming clearer, and different aspects of the rheology of dense suspensions and of dense shear-thickening suspensions are beginning to be untangled, we feel there is a need in the community to be more wary about the level of generality that can be drawn from specific systems. Unknown surface treatments made by a manufacturer on purchased particles, for example, can result in two apparently similar suspensions behaving in very different ways. In the same spirit, choices made on how to model different interactions in a simulation code may lead to behaviors that do not necessarily reflect how the modeled interaction influences the ``real-world'' suspension.

    \item Shear thickening in dense, non-Brownian suspensions seems to be viewed as a phenomenon in which the unexpected part is the thickening, and the thickened and shear-jammed states. We consider that it is not so, and consequently propose to shift perspective. The Wyart and Cates and constraints-based model both postulate that shear thickening stems from stress-activated constraints or frictional contacts. The short-range repulsive force is critical to achieve this unconstrained--constrained transition. It is the existence of this repulsive force between the particles that allows the suspension to be able to evolve onto two different rheological branches, and this ability is what is specific to shear-thickening suspensions. Both limit cases, the fully unconstrained state and the fully constrained state, have their own particular rheological behavior. In particular, the rheology of the fully constrained state is ``simply'' that of frictional granular suspensions~\cite{Lemaire_2023} (keeping in mind that it needs high enough stresses to exist). It is thus not surprising that the fully constrained state behaves as a dense frictional suspension: that is exactly what it is. Which means it is also not surprising that above the stress-dependent frictional jamming packing fraction ($\phi_c\left(\tau/\tau_R\right)$ as defined in Sec.~\ref{sec:WCmodel}), the suspension is shear-jammed in its fully constrained state. What is specific to shear-thickening suspensions is (i) that they can, at low stresses, relax back to an unconstrained state which can flow, and (ii) that they can exist in transitory states that are neither fully unconstrained nor fully constrained: the CST, pure-DST, and DST-SJ states.

    \item The previous point warrants a word of caution on using experimental results obtained on the thickened state of a shear-thickening suspension as results that would be general to dense frictional suspensions. It is not always easy to determine wether the suspension is indeed in its fully constrained state, and there might be some regions of the $\tau$--$\phi$ state diagram which we do not yet understand well enough. In line with the first point of this section, we urge the reader to be careful about hasty generalizations.

    \item Although a unified picture is emerging for dense suspensions and dry granular materials, a note of caution is warranted. In both systems, flow behavior is governed by the frictional contact network. In dry granular materials, a percolating frictional network is directly associated with the onset of shear jamming~\cite{Bi_2011, Ren_2013, Otsuki_2011,Zhao_2019}. In contrast, in dense shear-thickening suspensions, percolation occurs near the pure-DST regime~\cite{Boromand_2018, Sedes_2020, Naald_2024}, which lies close to, but still below, the SJ threshold. Likewise, while rigidity in dry granular materials aligns with the shear jamming point~\cite{Henkes_2016}, the emergence of a rigid frictional network in shear-thickening suspensions appears at or just before the packing fraction $\phi^*$ where pure DST is first observed~\cite{Naald_2024}.

    \item Most of the topological and geometrical approaches are limited to sliding constraints and two-dimensional settings only. As an example, the pebble game method to study the minimal rigidity is only applicable to sliding constraints (available for frictionless and frictional particle contacts) and does not have a three-dimensional counterpart. Real-world suspensions are three-dimensional. Thus, to make meaningful predictions in the real world, efforts are needed to be extended to 3D. Additionally, given that traditional discrete particle simulations are computationally expensive, large system sizes close to jamming are computationally intractable. In this spirit, machine learning is just starting to be a promising avenue, which needs more attention~\cite{Mandal_2022,Aminimajd_2025, Aminimajd_2025a}.

\end{enumerate}

\section{Acknowledgments}

The authors thank everyone who participated in the ``The Plot Thickens'' seminar series, as a speaker or as an attendee. This work is in large part the result of the many interesting discussions held during this monthly gathering. The authors also thank the masked geometer for their punny contribution. A. S. acknowledges the Case Western Reserve University for start-up funding.


%\nocite{*} % in order to check the references that are unused and remove them from the bib file

%\bibliographystyle{plain}
\bibliography{DST_Rheol_Acta}% common bib file

\end{document}